\documentclass[a4paper, doc]{apa6}
\usepackage{tikz}
\usetikzlibrary{backgrounds}
\usepackage{bbm}
\usepackage{apacite}
\usepackage{natbib}
\usepackage{times}
\usepackage{amssymb,amsmath,amsfonts}
\usepackage[hidelinks]{hyperref}
\usepackage{amsmath}
\usepackage{subfig}
\usepackage{graphicx}
\graphicspath{{Figures/}}
\usepackage{placeins}
\usepackage{booktabs}
\usepackage{rotating}
\usepackage[english]{babel}
\usepackage{epstopdf}
\usepackage{multirow}
\usepackage{array}
\usepackage{mathtools}
\usepackage[document]{ragged2e}
\usepackage{threeparttable}
\usepackage{caption}
\usepackage{color}
\usepackage[normalem]{ulem}
\usepackage{float}

\DeclareMathOperator{\param}{\boldsymbol{\theta}}
\def\btheta{\boldsymbol{\theta}}

\title{{\Large{\textbf{Multilevel latent class analysis with covariates:
Analysis of cross-national citizenship norms with a two-stage approach}}}}
\shorttitle{}
\fourauthors{Roberto Di Mari}{Zsuzsa Bakk}{Jennifer Oser}{Jouni Kuha}
\fouraffiliations{Department of Economics and Business, University of Catania, Italy}{Department of Methodology and Statistics, Leiden University, The Netherlands}{Department of Politics and Government, Ben-Gurion University, Israel}{Department of Statistics, London School of Economics and Political Science, London, UK}
\authornote{Corresponding author: Roberto Di Mari\\ Address: Corso Italia, 55, 95128, Catania, Italy.\\ Email: roberto.dimari@unict.it} 

\abstract{
	This paper focuses on the substantive application of multilevel LCA to the evolution of citizenship norms in a diverse array of democratic countries. 
	To do so, we present a two-stage approach to fit multilevel latent class models: in the first stage (measurement model construction), unconditional class enumeration is done separately on both low and high level latent variables, estimating only a part of the model at a time -- hence keeping the remaining part fixed -- and then updating the full measurement model; in the second stage (structural model construction), individual and/or group covariates are included in the model. 
	By separating the two parts -- first stage and second stage of model building -- the measurement model is stabilized and is allowed to be determined only by it's indicators.
	Moreover, this two-step approach makes the inclusion/exclusion of a covariate a relatively simple task to handle.
	Our proposal amends common practice in applied social science research, where simple (low--level) LCA is done to obtain a classification of low-level unit, and this is then related to (low-- and high--level) covariates simply including group fixed effects.
	Our analysis identifies latent classes that score either consistently high or consistently low on all measured items, along with two theoretically important classes that place distinctive emphasis on items related to engaged citizenship, and duty-based norms.}

\begin{document}
	\maketitle

	\section{Introduction}
	
	Latent class analysis (LCA) is an approach used extensively in the social sciences for creating a grouping based on a set of observed characteristics. Research using this analytic approach has addressed a variety of substantive topics, including for example  Swiss population opinion about support of pollution--reduction policies \citep{hill2001classification}, party performance in volatile systems \citep{mustillo2009modeling}, electoral incentives on congressional representation \citep{grimmer2013appropriators}, and profiles of citizenship norms and behaviors \citep{alvarez2017,hooghe2016,oser2017,oser2022}. 
	More methodological applications include modeling why support for extremist parties is underestimated in surveys \citep{breen}, and analysis of survey questions that include a ``don't know'' category \citep{feick}.
	Examples from other fields include, for instance, childhood socio--economic circumstances \citep{zhu2020}, profiles of drinking behavior in US adolescents \citep{chung2011,reboussin2010}, types of body mass index individuals \citep{brown2020}, and clusters of scientific journals based on bibliometric indicators \citep{bartolucci2015}. 
	
	In LCA one of the key assumptions is that sample units are independent after accounting for their class membership.
	The independence assumption is inadequate when data with a hierarchical structure are encountered -- as is often the case in the social sciences. Examples are survey respondents nested within regions/countries, children nested within families, and pupils within schools.
	In such cases, sound data modeling should take the nested data structure into account. 
	
	Multilevel extensions of LCA  and mixture models, which are able to take the nested data structure into account, are available (see, for instance, \citealp{asparouhov2008,vermunt.03,vermunt2008}). For example multilevel  LCA  for  nested  categorical  data, in  which  a  natural  grouping  is  observed, allows researchers to also  classify  the  groups (level 2 units) based  on  the  similarity  of  their  members (level--1 units). For nested data such as pupils in a school, a separate classification of both the pupils and the schools can be achieved (e.g., \citealp{fagginger,mutz2013}). While multilevel LCA is widespread, its use in political science and related fields is very limited; a recent example can be found in \cite{ruelens2020}. See also \cite{morselli2018,morselli2012}. 
	
	In practice, stepwise estimators are popular in LCA because they separate the estimation of the measurement and structural models. Substantive researchers see the estimation of these two parts as distinct steps, which are often even performed by different researchers at different time points \citep{bakk+kuha18,vermunt:10}. While in single level LCA the recent years have seen a breakthrough in the development of bias--adjusted three--step estimators \citep{asparouhov14b,Bolck:04,vermunt:10}, bias adjusted stepwise estimators are less available in the multilevel settings \citep{bakk2022}. 
		Furthermore, for models with distal outcomes, conceptually, adding the distal outcome at the same time as the indicators is even counterintuitive: the outcome defines the LC composition, while the latter should be predicted by the latent variable \citep{vermunt:10}. Having a readily available bias--adjusted stepwise estimator for multilevel LCA is therefore crucial.  
	
	Both asymptotic as well as finite--sample properties of such class of stepwise estimators have been widely studied, and are broadly understood. 
		However, bias--adjusted three--step estimators have one main drawback.
		Namely, they require computing classification error probabilities in their second step, which entails an explicit classification task. 
		With multilevel models, grouping occurs at two sequential layers.
		Therefore, an analogous step 2 in this context poses substantial problems, whilst classification is often not the goal of the analysis, but rather a juncture between estimation of measurement and structural models. 
	
	We hereby propose to extend a recently developed two-step estimator for LCA to the multilevel context \citep{bakk+kuha18}. Simulation results have shown that two--step estimators (well--known in the econometric literature) have very good finite sample properties.
		In addition, without an explicit classification step, they are more robust in complex scenarios, e.g. under differential item functioning \citep{dimaribakk}, or (mild) model misspecifications \citep{dimaribakkpunzo2020}. 
		In these and other challenging circumstances, the design of the two--step estimator can be easily adapted \citep{vermunt2021}. 
		
	We focus on the substantive application of multilevel LCA to the important topic of the evolution of citizenship norms in a diverse array of democratic countries.
	The rapid evolution of citizenship norms is considered one of the driving forces behind political processes in contemporary democracies.
	An important theoretical debate is whether more traditional, duty--based citizenship norms are gradually being replaced by more self--expressive and civically engaged values, variously referred to as ``engaged" \citep{coffe2010, dalton2008}, ``assertive" \citep{dalton2014, welzel2005}, or ``self--actualizing" \citep{bennett2008, bennett2012, shehata2016}. Although high--quality, cross-national data that is well--suited for latent class analysis is publicly available to inform this debate, guidance is needed regarding optimal multilevel modeling procedures to produce robust results.
	
	Common practice in applied social science research is to conduct simple LCA on the pooled data to obtain a classification of individual units.
	This is then related to individual and country level covariates, and the hierarchical data structure is taken into account by simply including country fixed effects in the regression.
	From the stepwise LC modeling literature, we know this approach to be wrong -- in the sense that it produces downward biased regression estimates.
	In addition, it can complicate estimation even with a moderate number of countries, and does not allow for a clustering at the country level -- which is useful from both methodological and application--oriented perspectives.
	
	Model building in multilevel LCA can be a challenging task for both conceptual and computational reasons.
	The first stage of this model building is the selection of the number of classes at the lower and group level.
	Related to this specific aspect, the standard approach is to jointly select the number of classes at both individual and group levels.
	This might make the preliminary stage of model selection extremely time consuming.
	\cite{Lukociene.10} suggest a stepwise approach to class selection, where in step 1 the lower level LCA model is defined in terms of number of classes.
	Then keeping the number of lower level classes fixed, the group level classes are estimated (step 2a).
	Next, a last step estimation at the lower level is re--iterated (step 2b) to re--adjust the number of lower level classes if necessary based on the final class solution at the higher level.
	
	This model building strategy has the advantage of reducing the computational burden of the simultaneous selection of the number of classes at both the lower and group levels which works equally well in practice (see, for example, \citealp{vermunt2008}).  
	Yet, \cite{Lukociene.10}'s multistep procedure requires estimating the full multilevel LC model several times at steps 2a and 2b of their procedure, which can still be time consuming if the choice of the number of low/high level classes is not informed by some a priori knowledge of the user, and/or is large.
	And even worse, with considerable sample size.
	
	In fact, in what follows we argue that step 2a can be carried out in a more efficient way by fixing the low--level parameters, and only estimating the high--level ones.
	This has the advantage of enhancing the computational efficiency of the whole selection procedure, and stepwise estimators of this kind are well--known to enhance computation stability even if model complexity increases (see, for instance, \citealp{bakk+kuha18,dimari2021,skrondal2012}).
	Then generalized information criteria with well--known statistical properties, commonly employed for model selection with pseudo-ML estimators, can be used to select the high--level number of classes (\citealp{takeuchi1976}; see also \citealp{bozdogan2000,lv2014}).
	Next step 2b, where the full measurement model is updated given the number of high level classes, becomes useful to stabilize the measurement model before the inclusion of covariates.
	
	Researchers gain important substantive insights from individual--level clustering, as well as from multilevel models that identify distinctive clusters in different group contexts (e.g., countries, schools).
	While individual and group level clustering are interesting \emph{per se}, there is often substantive interest in trying to predict individual and group clusters based on covariates.
	In the multilevel LCA context, when the number of candidate covariates is large the simultaneous estimator can be impractical: each time a predictor is added or removed, the full model must be re--estimated.
	In addition, algorithms for simultaneous estimation of models for nested data -- like multilevel and Markov models -- can be numerically unstable and slow to converge \citep{bartolucci2015_3}. 
	
	In the LCA literature it is widely acknowledged that including class predictors in the stage of selection of the number of classes might alter class enumeration and meaning \citep{asparouhov2012auxiliary,vermunt:10}. 
	The general recommendation for individual--level LCA is to add covariates after the number of classes is fixed \citep{nylund_masyn}.
	Compared to selection  of the number of classes at the lower and group level, adding predictors of class membership is seen both conceptually and practically a second stage of model building.
	However, general recommendations for including predictors in multilevel LC models are not available.
	
This paper introduces a new stepwise technique to fill these gaps: (i) by amending \cite{Lukociene.10}'s approach in the phase of class selection.; (ii) by specifying a two--stage estimator for the inclusion of predictors of class membership, which extends the two--stage estimation approach for single level LCA to the multilevel setting \citep{bakk+kuha18}.
		Operatively, the estimator is articulated as follows.
First, the lower level LC model is estimated for an increasing number of classes, and the number of classes is chosen such that some information criterion is optimized alongside substantive considerations about the number of meaningful classes.
Second, while both the number of low level classes and the low level LC (lower level measurement model) parameters are kept fixed, the group level number of classes is selected only estimating the parameters of the equations for the group level measurement model.
Third, parameters of the lower level measurement model are re--estimated, keeping fixed the group level model, in order to adjust for possible misspecifications due to higher level grouping.
Fourth, the structural model is built by loading covariates  on the (low and high level) latent variables, keeping the measurement model's parameters fixed.  

Very recently, in a large simulation study, the robustness against violation of the local independence assumption for the two--stage estimator for multilevel LC models has been investigated \citep{bakk2022}. 
	When all model assumptions are met, the two--stage has similar properties to the one-stage estimator. 
	Where direct effects between the covariate and the items of the
	latent class model are present, the two estimators have analogous performances.
	Namely,  as the severity of the underlying violations increases -- i.e. measurement noninvariance or differential item functioning (DIF) -- ignoring them leads to bias with both approaches.
	In related contexts, the general two--step approach was found to be generally more robust to violations of model assumptions in different parts of the joint model \citep{dimaribakk,dimaribakkpunzo2020}. 
	In the current work, we show how the two--stage multilevel LCA, thereby specifically proposed for the context of modeling DIF, can be used in more general settings as well.

This approach has the advantage of making both stages of multilevel LCA conceptually sounder and practically simpler.
We retain \cite{Lukociene.10}'s idea of hierarchical class selection, which is known to work well, and we substitute one of their full information ML steps with a consistent estimator.
Indeed, stepwise estimators of this type are well known to be consistent \citep{gong1981pseudo}, and estimation of only a part of the model at each stage greatly reduces the computational burden and enhances stability of the algorithms that are commonly used to perform ML estimation \citep{dimaribakk,dimari2016}.

Regarding covariate selection, every time a covariate is added to or removed from the model, class definitions and the optimal number of classes can change.
By separating the two parts -- first stage and second stage of model building -- the measurement model is stabilized and is allowed to be determined only by indicators.
Moreover, this makes the inclusion/exclusion of a covariate a relatively simpler task to handle, presumably even by different researchers at different time points on (partially) not overlapping samples.

Our stepwise approach is motivated by the desire to analyze comparative data on adolescents' conceptions of good citizenship gathered by surveys conducted by the International Association for the Evaluation of Educational Achievement (IEA).
Prior studies on citizenship norms have used latent class analysis with country as a covariate to investigate the evolution of citizenship norms between 1999 and 2009 \citep{hooghe2015, oser2013}, and the determinants of good citizenship for all available countries in the 2009 sample \citep{hooghe2016}. 
Our analysis in this paper focuses on applying a multilevel LCA approach to all 24 countries surveyed in 2016, which includes advanced Western European and Scandinavian democracies often included in high--quality surveys (e.g., Belgium, Denmark, Finland); a number of South American countries (e.g., Chile, Colombia, Mexico) and Asian countries (e.g., Chinese Taipei, Hong Kong, Korea); as well as a few Eastern European and lesser studied countries in these sorts of cross--national surveys (e.g., Russia, Malta).

The diversity of this high--quality cross--national data is a tremendous resource that also requires careful modeling to distinguish between predictors at the individual--level versus the country--level in order to obtain robust findings for content experts on these topics.
Using our stepwise estimator, we analyze the IEA data, and provide substantive interpretation of the findings as well as a step--by--step guide for software implementation. 	

The paper proceeds as follows.
In Section \ref{sec:method} we present the modeling framework; the two--stage estimator is introduced and described in detail in Section \ref{sec:2stage}, and the IEA data application is presented in Section \ref{sec:application}.
We conclude with final remarks regarding the paper's contributions and topics for future research in Section \ref{sec:conclusion}.

\section{The multilevel latent class model}\label{sec:method}


\subsection{Definition of the model}

Consider the vector of responses $\mathbf{Y}_{ij}=(Y_{ij1},\dots,Y_{ijK})$, where
$Y_{ijk}$ denotes the response of individual $i$ in group $j$ on the $k$--th categorical
indicator variable, with $1\leq k\leq K$ and $1\leq j\leq J$, where $K$ denotes the number of categorical indicators and $J$ the number of level 2 units.
In addition, we let $n_j$ denote the number of level 1 units within the $j$--th level 2 unit, with $1\leq j \leq J$.
For simplicity of exposition, we focus on dichotomous indicators.

LC analysis  assumes that respondents belong to one of the $T$ categories (``latent classes'') of an underlying categorical latent variable $X$ which affects the responses \citep{goodman:74,hagenaars:90,mccutcheon:87}. 
The model for $\mathbf{Y}_{ij}$
can then be specified as
\begin{equation}
	P(\mathbf{Y}_{ij})=\sum\limits_{t=1}^{T}P(X_{ij}=t)P(\mathbf{Y}_{ij}|X_{ij}=t),
	\label{eq:step1-marginal}
\end{equation}%
where the weight $P(X_{ij}=t)$ is the probability of person $i$ in group $j$ to belong to 
latent class $t$. The term $P(\mathbf{Y}_{ij}|X=t)$ is the class--specific
probability of observing a pattern of responses given that a person belongs to class $t$. 
Furthermore we
make the ``local independence'' assumption that the $K$ indicator
variables are independent within the latent classes, leading to
\begin{equation}
	P(\mathbf{Y}_{ij})=\sum_{t=1}^{T}P(X_{ij}=t)\prod%
	\limits_{k=1}^{K}P(Y_{ijk}|X_{ij}=t).
	\label{eq:step1-conditional}
\end{equation}%

Note that the general definition in Equation \eqref{eq:step1-marginal} applies to both the standard and multilevel LC model.
To be able to distinguish the simple and multilevel LC model we must define the model in terms of logit equations.
In the simple LC model

\begin{equation}
	P(X_{ij}=t)=\frac{\exp({\gamma_t})}{1 + \sum_{s=2}^T\exp(\gamma_{s})},
	\label{eq:logit_simpleLC}
\end{equation}%

for $1 < t \leq T$ -- where we have taken the first class as reference -- and

\begin{equation}
	P(Y_{ijk}\vert X_{ij}=t)=\frac{\exp({\beta_t^k})}{1 + \exp({\beta_t^k})},
	\label{eq:logitsimple_conditional}
\end{equation}%
for $k=1,\dots,K$.
In the simple LC model the parameters $\gamma$ and $\beta$ do not have the subscript $j$, thus assuming the clustering is independent of the group level classes. 

Extending the simple LC model to account for the multilevel data structure is possible by allowing the parametrizations  \eqref{eq:logit_simpleLC} and \eqref{eq:logitsimple_conditional} to take the grouping into account as follows

\begin{equation}
	P(X_{ij}=t)=\frac{\exp({\gamma_{tj}})}{1 + \sum_{s=2}^T\exp({\gamma_{sj}})}
			\label{eq:logit_MLLC}
		\end{equation}%
		
		again for $1 < t \leq T$ and
		
		\begin{equation}
			P(Y_{ijk}\vert X_{ij}=t)=\frac{\exp({\beta_{tj}^k})}{1 + \exp({\beta_{tj}^k})}.
			\label{eq:logitML_conditional}
		\end{equation}%
		for $k=1,\dots,K$.
		
		This is the most general formulation that is equivalent to an unrestricted multi--group LC model.
		In most applications however a more restricted version is used \citep{Lukociene.10,vermunt.03}  
		that assumes that item--conditional probabilities do not depend on the level 2 unit -- as in Equation \eqref{eq:logitsimple_conditional}.
		This is the restriction we will apply also in the data analysis in the current paper. 
		
		\subsection{Parametrizations of group--specific effects}
		The multilevel LC model can be seen as a random coefficients logistic regression model (see, for instance \citealp{agresti2000}), where the dependent variable is unobserved and has several observed indicators \citep{vermunt.03}. 
		An  example of such a model can be a LC model of lower level units of students, predicted by  several indicators of socio--economic status (e.g., parents education, income level) that are nested in higher level units of schools -– where clustering is identified at the school level that can be predicted by additional higher--level variables (e.g., average value per square meter of apartments in the area).  
		As with random coefficient logistic regression models, different ways of parametrizing multilevel LC models exist in the literature (see also \citealp{finch2014}), mainly falling into the \emph{parametric} and \emph{nonparametric} approaches. 
		
		In the parametric approach, group--specific effects are assumed to arise from a certain continuous distribution -- typically Gaussian.
		With $T$ latent classes, at least two alternatives are available.
		The first one requires to specify a ($T-1$) multidimensional distribution, say multivariate Gaussian.
		The second alternative, as suggested by \cite{vermunt.03}, is to work with the following set of equations
		\begin{equation}\label{eq:REequation}
			\gamma_{tj} = \gamma_{t} + \tau_{t} u_j,
		\end{equation}
		with $u_j \sim N(0,1)$, for $j=1,\dots,J$ and $t=2,\dots,T$.
		As before, the first class is taken as reference and the related parameters ($\gamma_1$ and $\tau_1$) are set to zero.
		Intuitively, Equation \eqref{eq:REequation} specifies random components sharing the same random effect $u_j$ across $t$'s, that is then scaled within each $t$ by the unknown $\tau_t$  (see also \citealp{vermunt2008}).

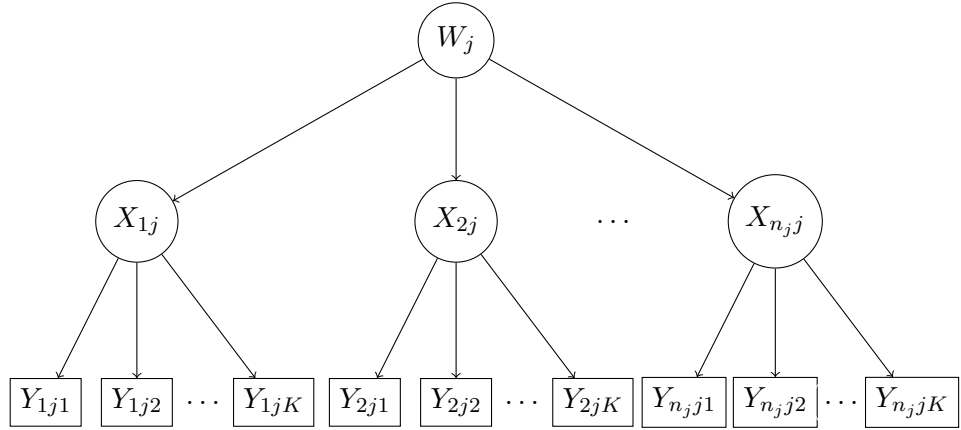
\begin{figure}[!t]
			\centering
			\begin{tikzpicture}[scale=0.6]
				\tikzstyle{every node}=[draw,remember picture] \node[circle] (v1) at (10,6)
				{$W_{j}$};
				\node[circle]
				(v3) at (3,2) {$X_{1j}$}; \node[circle] (v4) at (10,2)
				{$X_{2j}$}; \node[circle] (v5) at (17,2) {$X_{n_{j} j}$};
				\node[circle, draw=white] (v101) at (13.5,2) {$\dots$};
				
				\node[rectangle]
				(v61) at (1,-2) {$Y_{1j1}$}; \node[rectangle]
				(v62) at (3,-2) {$Y_{1j2}$}; \node[circle, draw=white]
				(v622) at (4.5,-2) {$\dots$}; \node[rectangle]
				(v63) at (6,-2) {$Y_{1jK}$};
				\node[rectangle] (v71) at (8,-2) {$Y_{2j1}$}; \node[rectangle] (v72) at (10,-2) {$Y_{2j2}$};
				\node[circle, draw=white] (v722) at (11.5,-2) {$\dots$};
				\node[rectangle] (v73) at (13,-2) {$Y_{2jK}$};
				
				\node[rectangle] (v81) at (15,-2) {$Y_{n_{j}j1}$};
				\node[rectangle] (v82) at (17,-2) {$Y_{n_{j}j2}$};
				\node[circle, draw=white] (v822) at (18.5,-2) {$\dots$};
				\node[rectangle] (v83) at (20,-2) {$Y_{n_{j}jK}$};
				
				\draw[->] (v1)
				-- (v3); \draw[->] (v1) -- (v4); \draw[->] (v1) -- (v5); \draw[->] (v3) -- (v61); \draw[->] (v3) -- (v62); \draw[->] (v3) -- (v63);
				\draw[->] (v4) -- (v71); \draw[->] (v4) -- (v72); \draw[->] (v4) -- (v73);
				\draw[->] (v5) -- (v81); \draw[->] (v5) -- (v82); \draw[->] (v5) -- (v83);
			\end{tikzpicture}
			\caption{Multilevel LC model. \label{fig:MLCM}}
			
		\end{figure}

		In the nonparametric approach, the continuous distribution is replaced with a discrete one \citep{aitkin1999,laird1978}, that is naturally assumed to be multinomial with $M$ mass points each with probability $P(W = m) = \pi_m$. Thus, we let $W$ be our multinomial group latent variable;
		by letting $W_j$ be the value of $W$ for group $j$, in the nonparametric approach the model for the (individual) latent class probabilities is specified as follows
		\begin{equation}\label{eq:NPREequation}
			P(X_{ij}=t \vert W_j = m) = \frac{\exp({\gamma_{tm}})}{1 + \sum_{s=2}^T\exp(\gamma_{sm})},
		\end{equation}
		where as before the first individual latent class is taken as reference and its parameters are set to zero for identification.
		Also the mixing probabilities $P(W = m)$ can be parametrized by means of logistic regressions as follows
		\begin{equation}\label{eq:pij}
			P(W = m) = \frac{\exp(\delta_{0m})}{1+\sum_{l=2}^M\exp(\delta_{0l})},
		\end{equation}
		where parameters for $m=1$ are set to zero for identification and the related class is set as reference. 
		The path diagram of the resulting multilevel latent class model is given in Figure \ref{fig:MLCM}. Note that $W_j$ affects the responses only through the low level LC variable – and not directly -- as in the restricted multi--group LC model.
		
		Such an approach, which this paper focuses on, allows the user 1) to work with weaker assumptions, and 2) to reduce the computational burden \citep{vermunt2001nonparametric} with respect to the parametric approach.
		In addition, the nonparametric approach has the byproduct of naturally providing a classification of the units into a seemingly small number of classes \citep{aitkin1999}.
		Under the nonparametric approach, the total number of free parameters to be estimated is $M-1 + M(T-1) + KT$.
		
		\subsection{Extending the model with covariates}

		Level 1 and level 2 covariates can be included to predict individual as well as group class membership.
		Denoting one level 2 covariate by $Z_{1j}$ and a level 1 covariate by $Z_{2ij}$, the model for $\mathbf{Y}_{ij} \vert \mathbf{Z}_j$, with $\mathbf{Z}_j = (Z_{1j}, Z_{2ij})^{\prime}$, can be specified as follows
		
		\begin{equation}\label{eq:condMLC}
			P(\mathbf{Y}_{ij} \vert \mathbf{Z}_j ) = \sum_{m=1}^M P(W_j = m \vert Z_{1j}) \sum_{t=1}^T P(X_{ij} = t \vert W_j = m, Z_{1j},Z_{2ij}) \prod_{k=1}^{K} P(Y_{ijk}|X_{ij} = t),
		\end{equation} 
		
		where the multinomial logistic regression for $X_{ij}$ with a random intercept now is as follows
		
		\begin{equation}
			P(X_{ij}=t \vert Z_{1j},Z_{2ij},W_j = m )=\frac{\exp({\gamma_{0tm}}+ \gamma_{1t}Z_{1j}+ \gamma_{2t}Z_{2ij})}{\sum_{s=2}^T \exp({\gamma_{0sm}}+ \gamma_{1s}Z_{1j}+ \gamma_{2s}Z_{2ij})}
					\label{eq:logit_covar_ml}
				\end{equation}%
				
				A random slope for the level 1 covariate can be obtained by replacing $\gamma_{2t}$ by $\gamma_{2mt}$, for $1< t \leq T$. 
				The multinomial logistic regressions for class membership of group $j$ $P(W_j = m)$ is specified now as 
				\begin{equation}\label{eq:logitWcovar}
					P(W_j = m \vert Z_{1j}) = \frac{\exp(\delta_{0m} + \delta_{1m} Z_{1j})}{1 + \sum_{l=2}^L \exp(\delta_{0l} + \delta_{1l} Z_{1j})}.
				\end{equation}
				
				Assuming a sample of $J$ groups is observed -- each with $n_j$ individual units, for $j=1,\dots,J$ -- the (log) likelihood function of can be specified as follows 
				\begin{equation}\label{eq:ll_mulLCA}
					\log L(\mathbf{\theta}) = \sum_{j=1}^{J} \log P(\mathbf{Y}_{j} \vert \mathbf{Z}_j),
				\end{equation}
				where $\mathbf{\theta}$ denotes the vector of all model parameters.
				The function \eqref{eq:ll_mulLCA} is maximized with respect to the unknown $\mathbf{\theta}$ to find ML estimates.
				This is typically accomplished by means of iterative procedures -- like the EM or Newton--type algorithms, or a combination of both. 
				
				However in practice, the choice of the number of latent classes on level 1 and level 2 is not so obvious.
				A generally used recommendation is to use a stepwise procedure for model selection \citep{Lukociene.10}, by first fitting the LC model at the level 1 -- as defined in Equations \ref{eq:logit_simpleLC} and \ref{eq:logitsimple_conditional}.
				Once the number of classes at the lower level is selected, this number is held fixed and the number of classes is estimated at the group level.
				A general recommendation is that, once the group level classes are selected, these are kept fixed, and model selection is re--iterated at the lower level one more time before adding covariates.
				In the stage of adding covariates the number of classes should be fixed, also to be in line with general recommendations for LCA with covariates \citep{masyn:dif}. 
				
				Using the procedure recommended by \cite{Lukociene.10}, the model is re--estimated to select the number of group level classes \citep{vermunt.03}.
				Also in this case, to add the covariates, while the number of classes is fixed the full model will be re--estimated again.
				Given the complexity of such multilevel models, 1) estimating the full model multiple times can be time consuming, and 2) misspecifications in a part of the model may also destabilize parameters in other parts of the model \citep{vermunt:10}.
				
				\section{Stepwise estimation of the multilevel LC model parameters} \label{sec:2stage}
\begin{figure}[!t]
					\centering
					\subfloat[\normalsize Step 1: simple latent class model on the pooled observations -- multilevel structure of the data is not taken into account. \label{fig:step1}]{
						\centering
						\begin{tikzpicture}[scale=0.6]
							\tikzstyle{every node}=[draw,remember picture] \node[circle] (v1) at (10,4)
							{$X$};
							
							\node[rectangle]
							(v61) at (1,-2) {$Y_{111}$}; \node[rectangle]
							(v62) at (3,-2) {$Y_{112}$}; \node[circle, draw=white]
							(v622) at (4.5,-2) {$\dots$}; \node[rectangle]
							(v63) at (6,-2) {$Y_{11K}$};
							\node[rectangle] (v71) at (8,-2) {$Y_{n_{j}j1}$}; \node[rectangle] (v72) at (10,-2) {$Y_{n_{j}j2}$};
							\node[circle, draw=white] (v722) at (11.5,-2) {$\dots$};
							\node[rectangle] (v73) at (13,-2) {$Y_{n_{j}jK}$};
							
							\node[rectangle] (v81) at (15,-2) {$Y_{NJ1}$};
							\node[rectangle] (v82) at (17,-2) {$Y_{NJ2}$};
							\node[circle, draw=white] (v822) at (18.5,-2) {$\dots$};
							\node[rectangle] (v83) at (20,-2) {$Y_{NJK}$};
							
							\draw[->] (v1) -- (v61); \draw[->] (v1) -- (v62); \draw[->] (v1) -- (v63);
							\draw[->] (v1) -- (v71); \draw[->] (v1) -- (v72); \draw[->] (v1) -- (v73);
							\draw[->] (v1) -- (v81); \draw[->] (v1) -- (v82); \draw[->] (v1) -- (v83);
						\end{tikzpicture}
					}\qquad
					\centering
					\subfloat[\normalsize Step 2.a: multilevel latent class model with measurement model kept fixed at Step 1's values. \label{fig:step2}]{
						\centering
						\begin{tikzpicture}[scale=0.6]
							\tikzstyle{every node}=[draw,remember picture] \node[circle] (v1) at (10,6)
							{$W_{j}$};
							\node[circle]
							(v3) at (3,2) {$X_{1j}$}; \node[circle] (v4) at (10,2)
							{$X_{2j}$}; \node[circle] (v5) at (17,2) {$X_{n_{j} j}$};
							\node[circle, draw=white] (v101) at (13.5,2) {$\dots$};
							
							\node[rectangle]
							(v61) at (1,-2) {$Y_{1j1}$}; \node[rectangle]
							(v62) at (3,-2) {$Y_{1j2}$}; \node[circle, draw=white]
							(v622) at (4.5,-2) {$\dots$}; \node[rectangle]
							(v63) at (6,-2) {$Y_{1jK}$};
							\node[rectangle] (v71) at (8,-2) {$Y_{2j1}$}; \node[rectangle] (v72) at (10,-2) {$Y_{2j2}$};
							\node[circle, draw=white] (v722) at (11.5,-2) {$\dots$};
							\node[rectangle] (v73) at (13,-2) {$Y_{2jK}$};
							
							\node[rectangle] (v81) at (15,-2) {$Y_{n_{j}j1}$};
							\node[rectangle] (v82) at (17,-2) {$Y_{n_{j}j2}$};
							\node[circle, draw=white] (v822) at (18.5,-2) {$\dots$};
							\node[rectangle] (v83) at (20,-2) {$Y_{n_{j}jK}$};
							
							\draw[->] (v1)
							-- (v3); \draw[->] (v1) -- (v4); \draw[->] (v1) -- (v5); \draw[->] (v3) -- (v61) node [draw=none, fill=none,midway, above, sloped] (TextNode) {fixed}; \draw[->] (v3) -- (v62) node [draw=none, fill=none,midway, above, sloped] (TextNode) {fixed}; \draw[->] (v3) -- (v63) node [draw=none, fill=none,midway, above, sloped] (TextNode) {fixed};
							\draw[->] (v4) -- (v71) node [draw=none, fill=none,midway, above, sloped] (TextNode) {fixed}; \draw[->] (v4) -- (v72) node [draw=none, fill=none,midway, above, sloped] (TextNode) {fixed}; \draw[->] (v4) -- (v73) node [draw=none, fill=none,midway, above, sloped] (TextNode) {fixed};
							\draw[->] (v5) -- (v81) node [draw=none, fill=none,midway, above, sloped] (TextNode) {fixed}; \draw[->] (v5) -- (v82) node [draw=none, fill=none,midway, above, sloped] (TextNode) {fixed}; \draw[->] (v5) -- (v83) node [draw=none, fill=none,midway, above, sloped] (TextNode) {fixed};
						\end{tikzpicture}
					}
					\vspace{0.5cm}
					\label{fig:step1_2}
					\caption{Steps 1 and 2.a of the proposed estimator to fit a multilevel latent class model. Note that Step 1 is equivalent to simple LCA on the pooled observations.}
				\end{figure}
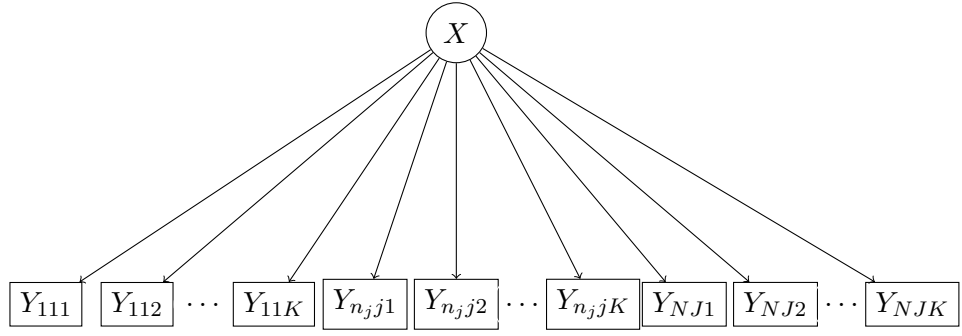
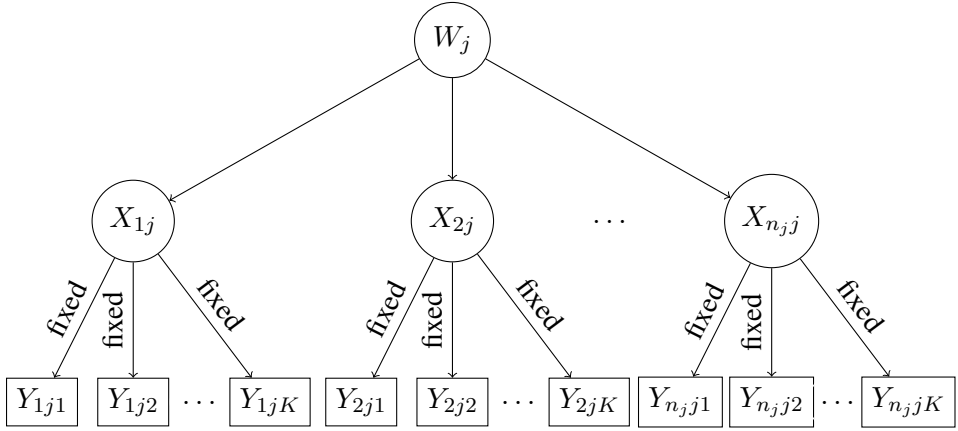

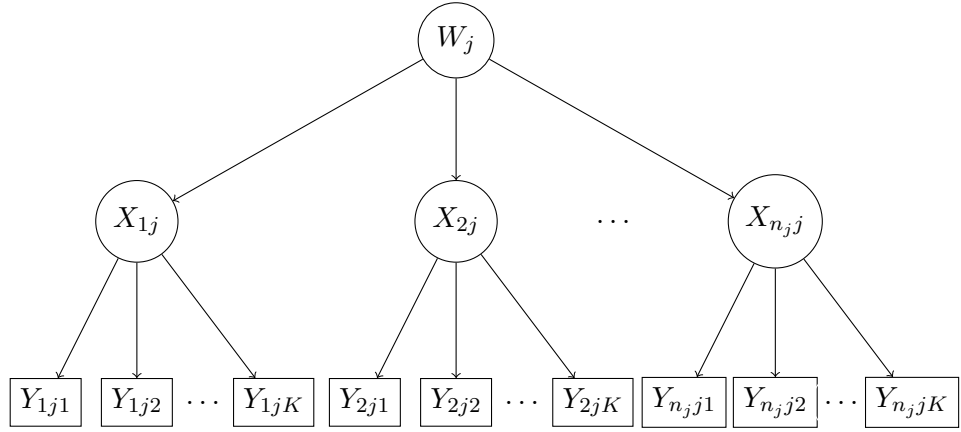
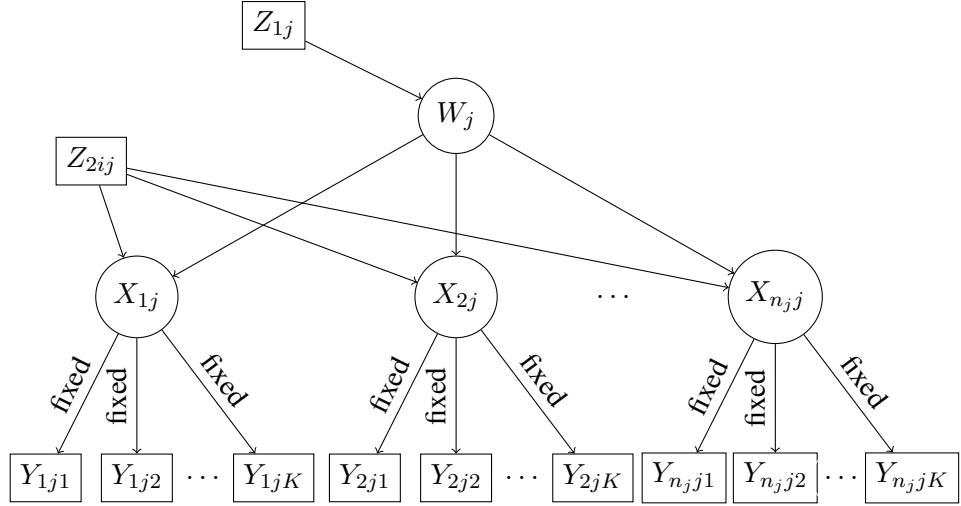
\begin{figure}[!t]
					\centering
					\subfloat[][\normalsize Step 2.b: measurement model is updated to account for possible interaction effects with high level parameters.  \label{fig:step3}]{
						\centering
						\begin{tikzpicture}[scale=0.6]
							\tikzstyle{every node}=[draw,remember picture] \node[circle] (v1) at (10,6)
							{$W_{j}$};
							\node[circle]
							(v3) at (3,2) {$X_{1j}$}; \node[circle] (v4) at (10,2)
							{$X_{2j}$}; \node[circle] (v5) at (17,2) {$X_{n_{j} j}$};
							\node[circle, draw=white] (v101) at (13.5,2) {$\dots$};
							
							\node[rectangle]
							(v61) at (1,-2) {$Y_{1j1}$}; \node[rectangle]
							(v62) at (3,-2) {$Y_{1j2}$}; \node[circle, draw=white]
							(v622) at (4.5,-2) {$\dots$}; \node[rectangle]
							(v63) at (6,-2) {$Y_{1jK}$};
							\node[rectangle] (v71) at (8,-2) {$Y_{2j1}$}; \node[rectangle] (v72) at (10,-2) {$Y_{2j2}$};
							\node[circle, draw=white] (v722) at (11.5,-2) {$\dots$};
							\node[rectangle] (v73) at (13,-2) {$Y_{2jK}$};
							
							\node[rectangle] (v81) at (15,-2) {$Y_{n_{j}j1}$};
							\node[rectangle] (v82) at (17,-2) {$Y_{n_{j}j2}$};
							\node[circle, draw=white] (v822) at (18.5,-2) {$\dots$};
							\node[rectangle] (v83) at (20,-2) {$Y_{n_{j}jK}$};
							
							\draw[->] (v1) -- (v3); \draw[->] (v1) -- (v4); \draw[->] (v1) -- (v5); \draw[->] (v3) -- (v61); 
							\draw[->] (v3) -- (v62); \draw[->] (v3) -- (v63);
							\draw[->] (v4) -- (v71); \draw[->] (v4) -- (v72); \draw[->] (v4) -- (v73);
							\draw[->] (v5) -- (v81); \draw[->] (v5) -- (v82); \draw[->] (v5) -- (v83);
						\end{tikzpicture}
					}\qquad
					\subfloat[][\normalsize Step 3: covariates $Z_{1j}$ and $Z_{2ij}$ are loaded respectively on $W_j$ and on $X_ij$, keeping measurement model parameters fixed. \label{fig:step4}]{
						\centering
						\begin{tikzpicture}[scale=0.6]
							\tikzstyle{every node}=[draw,remember picture] \node[circle] (v1) at (10,6)
							{$W_{j}$};
							\node[rectangle] (v11) at (6,8) {$Z_{1j}$};
							\node[circle]
							(v3) at (3,2) {$X_{1j}$}; \node[circle] (v4) at (10,2)
							{$X_{2j}$}; \node[circle] (v5) at (17,2) {$X_{n_{j} j}$};
							\node[circle, draw=white] (v101) at (13.5,2) {$\dots$};
							
							\node[rectangle] (v345) at (2,5) {$Z_{2ij}$};
							
							\node[rectangle]
							(v61) at (1,-2) {$Y_{1j1}$}; \node[rectangle]
							(v62) at (3,-2) {$Y_{1j2}$}; \node[circle, draw=white]
							(v622) at (4.5,-2) {$\dots$}; \node[rectangle]
							(v63) at (6,-2) {$Y_{1jK}$};
							\node[rectangle] (v71) at (8,-2) {$Y_{2j1}$}; \node[rectangle] (v72) at (10,-2) {$Y_{2j2}$};
							\node[circle, draw=white] (v722) at (11.5,-2) {$\dots$};
							\node[rectangle] (v73) at (13,-2) {$Y_{2jK}$};
							
							\node[rectangle] (v81) at (15,-2) {$Y_{n_{j}j1}$};
							\node[rectangle] (v82) at (17,-2) {$Y_{n_{j}j2}$};
							\node[circle, draw=white] (v822) at (18.5,-2) {$\dots$};
							\node[rectangle] (v83) at (20,-2) {$Y_{n_{j}jK}$};
							
							\draw[->] (v11) -- (v1);
							\draw[->] (v1)
							-- (v3); \draw[->] (v1) -- (v4); \draw[->] (v1) -- (v5);
							\draw[->] (v345) -- (v3); \draw[->] (v345) -- (v4);
							\draw[->] (v345) -- (v5);
							\draw[->] (v3) -- (v61) node [draw=none, fill=none,midway, above, sloped] (TextNode) {fixed}; \draw[->] (v3) -- (v62) node [draw=none, fill=none,midway, above, sloped] (TextNode) {fixed}; \draw[->] (v3) -- (v63) node [draw=none, fill=none,midway, above, sloped] (TextNode) {fixed};
							\draw[->] (v4) -- (v71) node [draw=none, fill=none,midway, above, sloped] (TextNode) {fixed}; \draw[->] (v4) -- (v72) node [draw=none, fill=none,midway, above, sloped] (TextNode) {fixed}; \draw[->] (v4) -- (v73) node [draw=none, fill=none,midway, above, sloped] (TextNode) {fixed};
							\draw[->] (v5) -- (v81) node [draw=none, fill=none,midway, above, sloped] (TextNode) {fixed}; \draw[->] (v5) -- (v82) node [draw=none, fill=none,midway, above, sloped] (TextNode) {fixed}; \draw[->] (v5) -- (v83) node [draw=none, fill=none,midway, above, sloped] (TextNode) {fixed};
						\end{tikzpicture}
					}
					\label{fig:steps3_4}
					\caption{Steps 2.b and 3 of the proposed estimator to fit a multilevel latent class model.}
				\end{figure}

				We propose an efficient stepwise estimator within a two--stage model building setup.
				Hereafter we outline the main idea of our two--stage approach, and give details of each step afterwards.
				\begin{enumerate}
					\item[A)] First stage: unconditional multilevel LC model building (measurement model construction).
					\begin{enumerate}
						\item[Step 1:] a simple latent class model is fitted on the pooled observations -- multilevel structure of the data is not taken into account (Figure \ref{fig:step1}).
						\item[Step 2.a:] the group level measurement model parameters of the multilevel LC model are estimated keeping the lower level measurement model parameters fixed at their first step values (Figure \ref{fig:step2}).
						\item[Step 2.b:] the full measurement model is re--estimated to account for possible interaction effects (Figure \ref{fig:step3}) between low and high level measurement parts
						This step is only necessary if the number of low and high level classes is not known a priori.
					\end{enumerate}
					\item[B)] Second stage: inclusion of covariates in the model (structural model construction). 
					\begin{enumerate}
						\item[Step 3:] covariates are included in the model (Figure \ref{fig:step4}) and only the structural model parameters are estimated keeping the measurement model fixed. 
					\end{enumerate}
				\end{enumerate}
				
				\subsection{Stage 1. Measurement model building}
				\subsubsection{Step 1: Simple LC model}
				In this step (Figure \ref{fig:step1}) a simple LC model is estimated on the pooled data $T_{\text{max}}$ times, where $T_{\text{max}}$ is pre--specified by the user.
				We let $\mathbf{\theta}_1^{(T)} = (\beta_2^1,\dots,\beta_T^1,\dots,\beta_2^K,\dots,\beta_T^K)^{\prime}$ for each choice of $T=1,\dots,T_{\max}$.
				Under the
				parametrizations \eqref{eq:logit_simpleLC} and \eqref{eq:logitsimple_conditional}, and a sample of $N$ observations -- where $N=\sum_{j=1}^J n_j$ -- the (log) likelihood function of the first step model can be specified as follows 
				\begin{equation}\label{eq:ll_LCA}
					\log L (\mathbf{\theta}_1) = \sum_{i=1}^{N} \log P(\mathbf{Y}_{ij}), 
				\end{equation}
				which we maximize in order to find the ML estimates of the LC model parameters, that we call $\widehat{\mathbf{\theta}}_1^{(T)} = (\widehat{\beta}_2^1,\dots,\widehat{\beta}_T^1,\dots,\widehat{\beta}_2^K,\dots,\widehat{\beta}_T^K)^{\prime}$.
				Then, the optimal number of classes $T^{*}$ is selected such that $T^{*} = \min\limits_{T=1,\dots,T_{\text{max}}} I(T)$ -- where $I(T)$ is some information criterion like AIC or BIC -- along with the corresponding ML estimates $\widehat{\btheta}_1$ from which we have suppressed the superscript $^{(T)}$ for simplicity of notation.
				
				\subsubsection{Step 2.a: Multilevel LC model}
				In this step (Figure \ref{fig:step2}), the group level measurement model parameters of the multilevel LC model are estimated keeping measurement model parameters at the lower level fixed at $\widehat{\btheta}_1$. Similarly as for Step 1, this step has to be carried out $M_{\text{max}}$ times, where $M_{\text{max}}$ is a pre--specified by the user number of latent classes for $W$.
				We let $\mathbf{\btheta}_2^{(M)} = (\delta_2,\dots,\delta_M,\dots,\gamma_{021},\dots,\gamma_{0T1},\dots,\gamma_{0TM})^{\prime}$ for each choice of $M=1,\dots,M_{\max}$. Under the
				parametrizations  \eqref{eq:logitsimple_conditional}, \eqref{eq:NPREequation} and \eqref{eq:pij}, and a sample of $J$ groups -- each with $n_j$ individual units, for $j=1,\dots,J$ -- the (log) likelihood function of the step 2.a model can be written as follows
				\begin{equation}\label{eq:ll_step2a}
					\log L (\mathbf{\btheta}_2 \vert \btheta_1 = \widehat{\btheta}_1^{T^{*}}) = \sum_{j=1}^{J} \log P(\mathbf{Y}_{j}),
				\end{equation}
				where $\vert \btheta_1 = \widehat{\btheta}_1$ indicates that the measurement model parameters are kept fixed at $\widehat{\btheta}_1$.
				The function \eqref{eq:ll_step2a} is maximized with respect to the unknown $\mathbf{\btheta}_2$ to find ML estimates.
				Note that only $M-1$ parameters are estimated in this step, out of $M-1 + M(T-1) + KT$ of the full multilevel LC model.
				
				The fact that some of the model parameters are kept fixed in this pseudo--ML step brings in a certain degree of misspecification.
				Therefore we propose finding the optimal number of high--level groups $M$ by minimizing the Takeuchi Information Criterion (TIC, \citealp{takeuchi1976}). The TIC is defined as follows
				\begin{equation}\label{eq:TIC}
					\text{TIC} = -2 \log L (\widehat{\btheta}_2 \vert \btheta_1 = \widehat{\btheta}_1) + 2 tr(\widehat{\mathbf{H}}),
				\end{equation}
				where $\widehat{\mathbf{H}} = \widehat{\mathbf{F}}^{-1} \widehat{\mathbf{R}}$ is an estimate of the contrast matrix and $\widehat{\mathbf{F}}^{-1}$ and $\widehat{\mathbf{R}}$ are respectively the estimates of inverse Fisher information  matrix  in inner productor (Hessian form) and of the outer product form of the  Fisher information matrix.
				Note that $\widehat{\mathbf{H}}$ is the well--known Lagrange multiplier test statistic and $tr(\widehat{\mathbf{H}})$ incorporates the effects of model complexity and model misspecification in a single term.
				
On a general note, the use of TIC is indeed very broad, and covers diverse areas of statistics, including regression and generalized linear models \citep{lv2014}, as well as  latent variable models (\citealp{vrieze2012};  see also \citealp{ranalli2016}).
				
				\subsubsection{Step 2.b: Re--update the measurement model}
				In this step (Figure \ref{fig:step3}) the measurement model is re--estimated for a given number of high--level classes.
				We argue that such a step is necessary especially if the number of low--level classes is unknown, to avoid overestimating $T$ due to unmodeled interactions between low and high level variability or poor lower--level class separation \citep{Lukociene.10}.
				In those cases, with $M$ selected from the previous step, this step is carried out $T_{\text{max}}$ times.
				
				Given a sample of $J$ groups -- each with $n_j$ individual units, for $j=1,\dots,J$, under the parametrizations  \eqref{eq:logitsimple_conditional}, \eqref{eq:NPREequation} and \eqref{eq:pij}, the (log) likelihood function of the step 2.b model can be written as follows
				\begin{equation}\label{eq:ll_step2b}
					\log L (\mathbf{\theta}_1 , \theta_2 ) = \sum_{j=1}^{J} \log P(\mathbf{Y}_{j}).
				\end{equation}
				
				This specification, with respect to that of Step 1, now takes the multilevel structure of the data into account.
				
				\subsection{Stage 2. The structural model with covariates}
				\subsubsection{Step 3: including predictors for class memberships}
				
				As a next step the covariates can be added to the model (Figure \ref{fig:step4}).
				At this stage also a decision needs to be taken whether a stepwise approach is preferred (adding first lower level covariates, and after fixing those adding at the group level) or all covariates can be added in a single step. The benefit of the first option can be robustness, however no simulation or theoretical results are available. For sake of conciseness, we will present the simultaneous step 3 -- its split counterpart can be derived analogously.
				
				Let us define $\mathbf{\theta}_3 = (\gamma_{12},\dots,\gamma_{1T}, \gamma_{22},\dots,\gamma_{2T})'$.
				With the parametrizations specified in Equations \eqref{eq:logitsimple_conditional}, \ref{eq:logit_covar_ml} and \eqref{eq:logitWcovar}, the model log--likelihood can be written as follows
				\begin{equation}\label{eq:step3}
					\log L (\mathbf{\theta}_2, \mathbf{\theta}_3 \vert \theta_1 = \widehat{\theta}_1) = \sum_{j=1}^{J} \log P(\mathbf{Y}_{j} \vert \mathbf{Z}_j),
				\end{equation}
				
				which we maximize with respect to $\mathbf{\theta}_2$ and $\mathbf{\theta}_3$ keeping $\mathbf{\theta}_1$ fixed at its step 2.b values $\widehat{\theta}_1$.
				
				\subsection{Statistical properties of the two--stage estimator}
				
				Our two--stage estimator is an instance of pseudo maximum likelihood estimation \citep{gong1981pseudo}.
				Such estimators are consistent under very general regularity conditions (see, for instance, \citealp{gourieroux1995}).
				Using our notation, let the true parameter vector be $\param^{\star} = (\param_1^{\star}, \param_2^{\star},\param_3^{\star})^{\prime}$.
				Then our two--stage estimate $\widehat{\param}$ is consistent for $\param^{\star}$ if the ML of $\param$ is itself consistent for $\param^{\star}$ and, most importantly, if 1) $\param_1$ and $\param_2$ and $\param_3$ can vary independently of each other, and 2) $\widehat{\param}_1$ is consistent for $\param_1^{\star}$.
				These conditions are satisfied in our case: 1) is satisfied since $\param_2$ and $\param_3$ are independent from $\param_1$; 2) is satisfied as well, since the parameter estimates at step 1 is a ML estimate of a simple LC model.
				
				From standard theory we know that our two--stage estimator is less efficient than the full information (one--step) ML estimator.
				In order to correctly estimate the true variance of our estimator, we suggest to use non--parametric bootstrap. 
				
				Regarding model selection, simultaneous model selection on a LC multilevel model without covariates should always be considered as the statistical first best.
				However, although it is the gold standard, the simultaneous selection strategy becomes quickly unfeasible even for moderate $T$ and $M$.
				This is why hierarchical model selection has been proposed in the literature by \cite{Lukociene.10}.
				All their steps are full ML steps, whereas we suggest a shortcut in step 2a. In our step 2a only the high--level parameters are estimated, and the low level parameters from the previous step are kept fixed.
				This is an instance of pseudo ML estimation: such an approach has the advantage of enhancing convergence speed, and also enhances computation stability \citep{bakk+kuha18}, especially when a priori information on the number of high level classes is not available.
				Then, available information criteria for misspecified models (\citealp{takeuchi1976}; see also \citealp{bozdogan2000,lv2014}) can be used instead of standard AIC or BIC to correctly assess the number of high--level classes.
				
				\subsection{Model selection: an \textit{extended approach}}
				
				In the previous subsections, we have advocated the use of information criteria to select the number of classes. 
				There can be real--data problems where one specification undoubtedly dominates the others: in such cases, the final choice is unquestionable.
				However, in certain occasions ``there is no particular reason to choose a single best model according to some criterion. 
				Rather it makes more sense to \textit{deselect} models that are obviously poor, maintaining a subset for further consideration.
				Sometimes this subset might consist of a single model, but sometimes perhaps not" \citep{kadane2004}. 
				
Consequently, a wise use of information criteria should be complemented with (i) the evaluation of class separation, and, perhaps most importantly, (ii) the inspection of comparable solutions (in terms of information criteria and class separation measures) to locate the most substantively meaningful model configuration(s) (see, among others, \citealp{magidsonvermunt2004}).
				
				Several options are available to assess class separation in latent variable models. 
				An extensive review can be found in \cite{LG51technical}; see also \cite{farcomeni2021} for an example with nested data.
				The most extensively used measure in the social sciences is \cite{magidson1981}'s entropy--based R$^{2}$. 
				In the multilevel context, the latter can be meaningfully defined at both lower and higher levels - see, e.g., \citep{Lukociene.10}.
				
				More specifically, let 
				\begin{description}
					\item[o] $P(X_{ij}=t, W_j = m \vert \mathbf{Y}_{ij}, Z_{1j}, Z_{2ij})$ be the joint (bivariate) posterior probability that individual $i$, within country $j$, belongs to low--level class $t$, and high--level class $m$, given the observed responses, and possibly covariates; 
					\item[o] $P(W_j = m \vert \mathbf{Y}_{ij}, Z_{1j})$ be the posterior probability that country $j$ belongs to high--level class $m$, given the observed responses, and possibly covariates; 
					\item[o] $P(X_{ij}=t \vert \mathbf{Y}_{ij}, Z_{1j}, Z_{2ij}) = \sum_{m=1}^{M} P(X_{ij}=t, W_j = m \vert \mathbf{Y}_{ij}, Z_{1j}, Z_{2ij})$ be the marginal posterior probability that individual $i$, within country $j$, belongs to low--level class $t$, given the observed responses, and possibly covariates.
				\end{description}
				
				The first two quantities are already available within the Expectation step of the EM algorithm used for ML estimation of the second stage model (further technical details can be found in \citealp{vermunt.03}).
				The marginal posteriors $P(X_{ij}=t \vert \mathbf{Y}_{ij}, Z_{1j}, Z_{2ij})$, as noted above, can be obtained without effort from the joint posteriors, by summing over $W_j$.
				
				The entropy--based R$^2$ measures the proportional reduction of entropy when both responses and covariates are available, compared to the case where these are unknown. Its general formulation is as follows
				\begin{equation}\label{eq:entrR2}
					\text{R}^2_{\text{entr,low/high}} = \frac{E_{\text{TOT,low/high}} - \bar{E}_{\text{low/high}}}{E_{\text{TOT,low/high}}},
				\end{equation} 
				where the subscript ``low/high'' indicates that the general definition can be applied to both hidden layers, and ``$E$'' stands for entropy.
				
				At lower level, $E_{\text{TOT,low}} = \sum_{t=1}^T - \widehat{p}_{t} \log \widehat{p}_{t}$, where $\widehat{p}_{t} = \frac{1}{N} \sum_{j=1}^{J} \sum_{i=1}^{n_j} P(X_{ij}=t \vert \mathbf{Y}_{ij}, Z_{1j}, Z_{2ij})$, and $N$ is the total sample size; $\bar{E}_{\text{low}} = \frac{1}{N} \sum_{j=1}^J \sum_{i=1}^{n_j} - P(X_{ij}=t \vert \mathbf{Y}_{ij}, Z_{1j}, Z_{2ij}) \log \left\{ P(X_{ij}=t \vert \mathbf{Y}_{ij}, Z_{1j}, Z_{2ij}) \right\}.$
				
				At higher level, $E_{\text{TOT,high}} = \sum_{m=1}^{M} - \widehat{\omega} \log \widehat{\omega}$, where $\widehat{\omega} = \frac{1}{J} P(W_j = m \vert \mathbf{Y}_{ij}, Z_{1j})$, and $\bar{E}_{\text{high}} = \frac{1}{J} - P(W_j = m \vert \mathbf{Y}_{ij}, Z_{1j}) \log\left\{P(W_j = m \vert \mathbf{Y}_{ij}, Z_{1j}) \right\}$. 
				
				In order to compare and/or rank similar solutions, any decision process, based on information criteria, and on class separation measures, substantively benefits from a comparative evaluation of the key model features. 
				These include characteristics of the measurement, e.g. the lower and higher level different item--class profiles, as well as the distinct structural model estimates that are obtained for different choices of the number of groups. 
				Such an \textit{extended approach} is an important part of our empirical strategy, and, more in general, what we recommend using in multilevel latent class analysis with external predictors.

				\section{Multilevel LCA of cross--national citizenship norms}\label{sec:application}
				\subsection{Data description}
				As part of a comprehensive evaluation of education systems, the IEA conducted surveys in 1999, 2009 and 2016 in school classes of 14--year olds to investigate civic education with the same scientific rigor as the evaluation of more traditional educational skills of language and mathematics.
				
				The target population of the survey is defined as all students in Grade 8 (approximately 14--year old), provided that the average age of students in this grade was 13.5
					years or above at the time of the assessment.
					If, on average, students in Grade 8 were below
					13.5 year old, Grade 9 became the target population.
					The school samples were designed as stratified two--stage cluster samples; schools were randomly selected at the first stage with probability proportional to size, and intact classrooms were sampled at the second stage.
					
					Typically, each country aimed for a sample size of 150 schools, the total number of sampled students ranging between 3000 and 4500.
					The sample participation requirement for the survey
					was 85 percent participation of the selected schools and 85 percent of the selected students
					within the participating schools, or a weighted overall participation rate of 75 percent. Countries that did not meet the required response rate are excluded from the sample. 
					Additional details can be found in \cite{schulz2018IEA}.
				
				More in general, the IEA's survey design is comparable over time, and the involvement of educational authorities in the participating countries' school--based fieldwork ensures a high level of representativeness for a respondent--based survey.
				The resulting data are therefore not compromised by sampling challenges that researchers increasingly face in efforts to obtain representative survey data of adults through common approaches of random sampling through telephone, face--to--face interviews, or online panel samples.
				Due to this internationally comparable and representative sampling design, data from these surveys has been used to inform research on a variety of topics related to civic attitudes over time, ranging from attitudes towards migrants \citep{amnaa2010,munck2018} to attitudes on citizenship learning and education studies \citep{barber2012,knowles2018}, and tolerance and civic engagement \citep{sandoval2018,torney2002}. 
				
				The survey includes a comprehensive battery of questions on citizenship norms that can be validly compared across time and place: in 1999 in 28 countries \citep{schulz2004}, in 2009 in 38 countries \citep{schulz2011}, and in 2016 in 24 countries \citep{schulz2018}.
				The present study focuses on the third wave of the survey that was conducted in 2016. The full sample comprises 94603 units, but we focus on complete cases only -  87123 respondents, as about 8\% of the full sample has missing values on at least one indicator. More in detail, each country is measured from 1300 up to more than 7000 times by low--level units over 12 indicators (see Figure \ref{tab:country_sampsize}).
				This, according to recent recommendations \citep{park2018}, is a safe environment for multilevel LCA with the nonparametric approach, even with low class distinctness and high complexity of the latent structure – where this is defined by the number of latent clusters and classes at the higher and lower levels.
				
				\begin{table}[!h]
					\centering
					\begin{tabular}{l c}
						\hline
						\hline
						Country & sample size\\
						\cline{1-2}
						Belgium  &  2750  \\ 
						Bulgaria  &  2682  \\ 
						Chile  &  4753  \\ 
						Colombia  &  4992  \\ 
						Denmark  &  5692  \\ 
						Germany  &  1313  \\ 
						Dominican Republic  &  2779  \\ 
						Estonia  &  2770  \\ 
						Finland  &  3037  \\ 
						Hong Kong  &  2553  \\ 
						Croatia  &  3655  \\ 
						Italy  &  3274  \\ 
						Republic of Korea  &  2557  \\ 
						Lithuania  &  3422  \\ 
						Latvia  &  3000  \\ 
						Mexico  &  4987  \\ 
						Malta  &  3317  \\ 
						Netherlands  &  2692  \\ 
						Norway  &  5740  \\ 
						Peru  &  4713  \\ 
						Russia  &  7049  \\ 
						Slovenia  &  2664  \\ 
						Sweden  &  2828  \\ 
						Taiwan  &  3904  \\
						\hline \hline
					\end{tabular}
					\caption{Number of respondents per country of the third wave (2016) of the IEA survey used for the analysis.}
					\label{tab:country_sampsize}
				\end{table}

				Questions regarding citizenship norms in all three waves asked respondents to explain their understanding of what a good adult citizen is or does.
				The survey then lists a variety of activities for respondents to rate in terms of how important these activities are in order to be considered a good adult citizen. 
				The twelve items range from obeying the law and voting in elections, to protecting the environment and defending human rights.
				The items are presented in random order, so that respondents can indicate their own individual set of specific preferences.
				The survey question regarding citizenship norms is worded as follows: \citep{kohler2018} Q23 How important are the following behaviors for being a good adult citizens? Respondent options include ``very important"; ``quite important"; ``not very important"; and ``not important at all". The options ``very important" and ``quite important" are coded as 1, and ``not very important" and ``not important at all" are coded as 0.
				
				Citizenship norm items included in the analysis are as follows: Obeying the law (\textit{obey}), taking part in activities promoting human rights (\textit{rights}), participating in activities to benefit people in the community (\textit{local}), working hard (\textit{work}), taking part in activities to protect the environment (\textit{envir}), voting in every election (\textit{vote}), learning about the country’s history (\textit{history}), showing respect for government representatives (\textit{respect}), following the political issues in the newspaper on the radio or on tv (\textit{news}), participating in a peaceful protest against a law believed to be unjust (\textit{protest}), engaging in political discussions (\textit{discuss}), and joining a political party (\textit{party}).
				See appendix for full question wording for the citizenship norm indicators.  
				
				Covariates included in the analysis are customary determinants of citizenship norms from the literature at the individual--level of socio--economic measures and country--level measure of gross domestic product (GDP) per capita.
				For the individual--level covariates, prior studies have found gender and socio--economic characteristics to be important covariates with citizenship norms, with engaged adolescents more likely to be female and have greater socio--economic advantages.
				We follow standard practice of research on adolescents to operationalize socio--economic status through a proxy measure of the number of books at home, along with measures of respondents' educational expectations, and parental education \citep{andrew2011, hooghe2016}.
				See the appendix for additional information on covariate question wording and variable categories.
				
				Data preparation and model fit have been carried out with Latent GOLD 5.1 \citep{LG51technical} in combination with R\footnote{Data preparation, as well as estimation code and syntax files for replication are available from the corresponding author upon request.}.
				
				\subsection{Results}\label{sec:resappl}
				
				In this section we first report on the measurement model resulting from the first stage of the two--stage estimator. Then, we relate the low and high level latent classes to covariates and -- for sake of comparison -- we present the regression outputs for the simultaneous (one--stage) estimator and the computationally more efficient two--stage estimator.
				Subsequently, we compare both estimators for the multilevel latent class models to the results obtained with the na\"ive estimator -- which relates external variables to class membership through a simple multinomial regression of the estimated vector of class memberships on covariates, including country fixed effects. 
				
				We selected the number of classes based on the procedure described in Section \ref{sec:2stage}.
				A preliminary assessment of $T$ (Step 1) was done by fitting a simple LC model on the pooled data for $T=1,\dots,10$.
				As displayed by the model summaries in Table \ref{tab:modsum_step1}, the minimum BIC model is achieved at $T=10$.
				This is unfortunately only a ceiling effect, as by increasing $T$ up to 15 lowers the BIC further.
				Balancing BIC values with a class separation measure (the entropy--based $R^2$), a fairly good choice of $T$ is 5.
				
				We thus proceed to Step 2a, fixing $T=5$ and selecting $M$ between 1 and 4 (we do not go beyond 4 to avoid class uninterpretability and parameter proliferation).
				Model summaries are displayed in Table \ref{tab:modsum_step2a}. Together with TIC values, we also report the BIC with penalty term computed based on the high--level sample size following \cite{Lukociene.10}'s recommendation, and the standard BIC based on low--level sample size. The best model according to TIC -- and all other criteria -- is with $M=3$.
				
				Table \ref{tab:modsum_step2b} reports the summary information for model fits of Step 2b.
				In this case, the minimum BIC model is achieved at $T=10$, although the best solutions balancing model fit with parsimony and class separation are with T ranging between 4 and 6.
					Arguably, in terms of class interpretation, $T=4$ offers the most substantively meaningful model configuration (see Figure \ref{fig:Figbasic}).
					This is also consistent with previous research on prior waves of the same survey conducted in 1999 and 2009 \citep{hooghe2016,hooghe2015,oser2013}.
					The 2 new classes identified in the 5- and 6-class models offer more nuanced variations on the 4-class solution but are less theoretically coherent in relation to the literature, and tend more to mean scores on all measures.
				
				For the 5-class solution (see \ref{fig:Figbasic_cl5}), based on class sizes, the new class seems to emerge almost completely from taking the ``duty" class in the 4-class solution that was 25.58\% of the population, and splitting it in the 5-class solution into LC2 (duty, 16\%), and the new LC5, 12\%. 
					This can be identified as somewhat between engaged and duty - we could also call it `` mainstream'', since its values are around the mean prevalence of the indicators in the overall sample.
					The 6-class solution (see \ref{fig:Figbasic_cl6}) includes a new class - cluster 4 - that is akin to the 5-class solution's new ``mainstream'' group but with some elements of the ``maximalist'' group; and the additional ``mainstream'' class (cluster 6) has similar emphases, with particularly high scores on news and discuss. 
				
				Overall, we find that results from the simultaneous and two--stage estimators are very similar, which is to be expected in general when model assumptions are not violated.
				However we report that the simultaneous estimator is about 10 times slower than that of the two--stage estimator in terms of computing time.
				In addition, in line with the cited literature on stepwise estimators, we observed a general enhanced algorithmic stability and speed of convergence of the two--stage estimator (in terms of number of algorithm iterations) with respect to the simultaneous estimator.
				
				The LC response probabilities -- as after step 2b -- are reported in Table \ref{tab:lca_low} (right part)\footnote{Figure \ref{fig:Figbasic} gives a visual representation of the class profiles.}.
				The findings point to four main distinctive citizenship norms.
				Two of these latent classes are distinctively characterized by conditional probabilities that are consistently high (the ``maximal" group, 44\%) or consistently low (``subject" group, 4\%).
				The remaining two latent classes are characterized by distinct emphases of the type of indicators of citizenship they view as most important for good citizenship, and relate to two theoretically important concepts in the literature on citizenship norms.
				The ``engaged" group (27\%) emphasizes the types of norms most associated with more recently prevalent engaged attitudes (e.g., protecting the environment) while placing less emphasis on more traditional citizenship activities (e.g., party membership).
				Alternatively, the ``duty" class (26\%) places a high emphasis on the importance of voting, discussing politics, and party activity, but expresses relatively low interest in engaged norms of protecting human rights, and engaging in activities to benefit the local community. 
				
				\begin{table}[!h]
					\resizebox{\textwidth}{!}{\begin{tabular}{l c c c c c c c }
							\hline \hline
							&	HL class 1	&	HL class 2 	& HL class 3 	&	LL class 1	&	LL class 2	&	LL class 3	&	LL class 4	\\
							\cline{2-8}
							&	(25\%) & (50\%) & (25\%)  &	Maximal (44\%)	&	Engaged (27\%)	&	Subject (4\%)	&	Duty (26\%)	\\

							& 	& 		& 		&			&			&			&		\\
							\hline
							\textit{obey}	&	0.89	&	0.94	&	0.91	&	0.98	&	0.91	&	0.47	&	0.89	\\
							\textit{rights}	&	0.75	&	0.91	&	0.92	&	0.99	&	0.97	&	0.15	&	0.68	\\
							\textit{local}	&	0.72	&	0.88	&	0.88	&	0.97	&	0.92	&	0.12	&	0.64	\\
							\textit{work}	&	0.80	&	0.87	&	0.84	&	0.93	&	0.83	&	0.46	&	0.78	\\
							\textit{envir}	&	0.77	&	0.92	&	0.94	&	0.99	&	1.00	&	0.27	&	0.69	\\
							\textit{vote}	&	0.80	&	0.89	&	0.84	&	0.98	&	0.80	&	0.29	&	0.79	\\
							\textit{history}	&	0.73	&	0.88	&	0.88	&	0.96	&	0.91	&	0.28	&	0.64	\\
							\textit{respect}	&	0.77	&	0.84	&	0.81	&	0.89	&	0.81	&	0.24	&	0.77	\\
							\textit{news}	&	0.72	&	0.83	&	0.69	&	0.96	&	0.58	&	0.15	&	0.72	\\
							\textit{protest}	&	0.53	&	0.74	&	0.68	&	0.89	&	0.65	&	0.12	&	0.41	\\
							\textit{discuss}	&	0.38	&	0.55	&	0.34	&	0.76	&	0.12	&	0.03	&	0.34	\\
							\textit{party}	&	0.30	&	0.44	&	0.30	&	0.60	&	0.16	&	0.04	&	0.25	\\
							
							\cline{2-3}
							Maximal (LL  class 1)	&	0.22	&	0.62	&	0.31
							&	&		&		&	\\
							Engaged (LL class 2)	&	0.14	&	0.17	&	0.56
							&	&		&		&	\\
							Subject (LL class 3)	&	0.06	&	0.03	&	0.04
							&	&		&		&	\\
							Duty (LL class 4)	&	0.58	&	0.18	&	0.09
							&	&		&		&	\\
							\hline \hline
					\end{tabular}}
					\caption{The lower and higher level class latent class model averaged across all countries.}
					\label{tab:lca_low}
				\end{table}

				At the group class level, item probabilities are reported in Table \ref{tab:lca_low} (left part).
				The main distinction is between a group class with relatively low scores on all indicators (HL class 1, which is characterized by a predominance of more advanced democracies, e.g., Belgium, Denmark, Sweden); a group class with higher scores on all indicators (HL class 2, which is characterized by a predominance of least developed democracies, e.g., Croatia, Russia, Taiwan); and a group with mid--range scores that is characterized by mid--range levels of democratic development (HL class 3, Bulgaria, Chile, Colombia).
				These findings are consistent with prior related research (e.g., \citealp{hooghe2016}). Detailed country profiles are available in the appendix.

						In Table \ref{tab:regtable} we report results of the structural model parameters for the one--stage, two--stage and na\"ive estimators.
						Overall, we observe that parameter values are similar for all estimators.
						Also inference on the parameter values leads to the same conclusions in most cases, although the na\"ive estimator seems to show the well-documented tendency to have an inflated type--I error (\citealp{dimari2016,vermunt:10}); see, for instance, parameter and SE estimates for Education goal and Non--native language level.
						
						The regression findings are consistent with prior research on this topic.
						The engaged and maximal groups are more female, while the duty and subject groups are more male.
						When socio--economic distinctions are identified, the engaged norm is generally characterized by higher educational and socio--economic status measures than the other identified norms. 
						
						\begin{table}[!h]
							\resizebox{\textwidth}{!}{\begin{tabular}{l ccc ccc ccc}
									\hline \hline
									& \multicolumn{9}{c}{Higher level structural model}\\
									\cline{2-10}
									&	\multicolumn{2}{c}{Class 2}	&		& &	\multicolumn{2}{c}{Class 3}	 &	&		&	\\
									\cline{2-3} \cline{6-7}
									&one--stage	&	two--stage	&  & 	& one--stage	&	two--stage	&  &	 &	\\
									log GDP 		& -7.33$^{***}$ & -7.33$^{**}$		&	& 	& -4.67$^{*}$	& -4.67$^{**}$  &  &	 &	\\
									& (3.34) &	(3.31)		 	&	& 	& (2.73)	& (2.36) &  &  &  \\
									
									\midrule
									& \multicolumn{9}{c}{Lower level structural model}\\
									\cline{2-10}
									&	\multicolumn{3}{c}{Class 2 (Engaged)}	&	\multicolumn{3}{c}{Class 3 (Subject)}	&	\multicolumn{3}{c}{Class 4 (Duty)}\\
									&one--stage	&	two--stage	&	na\"ive	&	one--stage	&	two--stage	&	na\"ive	&	one--stage	&	two--stage	&	na\"ive	\\
									\cline{2-10}
									SES proxy &  -0.26$^{**}$ &  -0.25$^{**}$ &  0.05 &  -0.25$^{**}$ &  -0.24$^{*}$ &  -0.42$^{***}$ &  -0.12$^{**}$ &  -0.12$^{**}$ &  -0.15$^{***}$\\ 
									& (0.09) & (0.09) & (0.03) & (0.10) & (0.12) & (0.09) & (0.04) & (0.04) & (0.04)\\ 
									Female &  0.37$^{***}$ &  0.38$^{***}$ &  0.39$^{***}$ &  -0.46$^{**}$ &  -0.47$^{**}$ &  -0.52$^{***}$ &  -0.22 &  -0.2$^{*}$ &  -0.19$^{***}$\\ 
									& (0.11) & (0.10) & (0.05) & (0.15) & (0.15) & (0.12) & (0.14) & (0.10) & (0.06)\\ 
									Education goal &  0.14 &  0.16 &  0.12$^{*}$ &  -0.65$^{***}$ &  -0.64$^{***}$ &  -0.53$^{***}$ &  -0.35$^{***}$ &  -0.35$^{***}$ &  -0.34$^{***}$\\ 
									& (0.13) & (0.12) & (0.06) & (0.05) & (0.08) & (0.10) & (0.05) & (0.08) & (0.05)\\ 
									Mother education &  0.00 &  0.01 &  0.02 &  0.03 &  0.04 &  0.12 &  0.01 &  0.01 &  0.00\\ 
									& (0.07) & (0.08) & (0.04) & (0.04) & (0.06) & (0.10) & (0.03) & (0.04) & (0.04)\\ 
									Father education &  -0.10 &  -0.09 &  0.09$^{*}$ &  -0.07 &  -0.07 &  -0.15 &  -0.09$^{**}$ &  -0.09$^{*}$ &  -0.06\\ 
									& (0.08) & (0.09) & (0.04) & (0.08) & (0.08) & (0.09) & (0.03) & (0.04) & (0.04)\\ 
									Non-native language level &  -0.19 &  -0.22 &  0.01 &  0.14 &  0.15 &  0.63$^{*}$ &  0.26 &  0.27 &  0.20\\ 
									& (0.17) & (0.18) & (0.11) & (0.35) & (0.25) & (0.25) & (0.25) & (0.30) & (0.10)\\ 			
									\hline \hline
							\end{tabular}}
							\caption{Structural model parameter estimates for the simultaneous (one--stage) estimator, the two--stage estimator and the na\"ive estimator, with 4 lower level and 3 higher level classes. The number of books available at home is taken as socio--economic status (ses) proxy. First lower and upper level classes are taken as reference for the multinomial logistic regressions. Country log GDP is included also as low--level control for the na\"ive approach (not reported). Standard errors in parentheses (standard errors for the two--stage estimator are obtained with non parametric bootstrap on the 4th step). $^{***}$ $p$--value$<$0.01, $^{**}$ $p$--value$<$0.05, $^{*}$ $p$--value$<$0.1}
							\label{tab:regtable}
						\end{table}
						
						

						\FloatBarrier
						\section{Discussion}\label{sec:conclusion}
						
						In this paper we have proposed a two--stage estimator for multilevel LCA to model citizenship norms with the IEA data
						We have shown that the proposed estimator greatly simplifies model construction compared to the simultaneous estimator, although they both produce almost equal results in terms of structural model parameter estimates.
						This is to be expected if model assumptions are not violated.
						On a practical note, we report that the simultaneous estimator was about 10 times slower than that of the two--stage estimator in terms of computing time to produce parameter estimates
						More in general, we observed a general enhanced algorithmic stability in terms of convergence to stationary points with no boundary values, and speed of convergence, in terms of number of algorithm iterations, of the two--stage estimator  with respect to the simultaneous estimator, in line with the cited literature on stepwise estimators.
						
						In our empirical analysis we have also included results from the approach (na\"ive approach) that is current common practice in applied social science research.
						We report that we observed non-convergence (the maximum number of iterations is reached but the difference of two subsequent log-likelihood values was never smaller than or equal to 10$^{-8}$) and boundary parameter estimates due to the large number of country fixed effects that are included in the model.
						In addition, although the overall conclusions are the same, the na\"ive estimator seems to show the well-documented tendency to have an inflated type--I error (\citealp{dimari2016,vermunt:10}). 
						Our recommendation is to use the na\"ive approach only if the number of group--level units is small, and to generally rely on multilevel modeling.
						
						Consistent with prior research on citizenship norms, the findings identify several distinctive normative types.
						In addition to the theoretically relevant classes that emphasize engaged versus duty--based types of norms, the findings also identify classes that consistently score either high or low on all normative items.
						The socio--demographic covariates are consistent with prior research, confirming that the engaged norm adherents among adolescents tends to be more female, and characterized by higher socio--economic status.
						
						Whereby the two--stage estimator has been used with the IEA data, more generally this estimator is well suited for all data sets with a similar nested structure -- and possibly several candidate covariates.
						As a general recommendation based on our findings, although simultaneous estimators are theoretically to be preferred over stepwise estimators for their statistical properties, the two--stage estimator for LC multilevel models is a more efficient and stable alternative.
						An issue that is worth still further investigation using multilevel LCA (both one and two--step) is model selection and model selection criteria.
						While AIC and BIC are readily available in commercial software for multilevel models or, like the TIC, can easily be derived from available output, other more sophisticated model selection approaches such as the bootstrap likelihood--ratio test \cite{nylund2007} have not yet been extended to these family of models.
						
						Similar two--stage (or two--step) approaches have been already proposed in other contexts, like simple LC or latent Markov modeling.
						A multi-stage (hierarchical) class selection approach has been explored by \cite{Lukociene.10} in multilevel models, but the two--stage estimator is new in this context: multilevel data have their own specific features, which require appropriate tools.
						Note that, although longitudinal data can be viewed as nested data (individuals are high--level units, time points are low--level units), the modeling approaches are conceptually and practically different.
						
						%

						In this paper we implemented one of the two main ways we described as approaches for conducting step 2.b, namely we fix all parameters in $W$ and $X \mid W$ multinomial logistic regressions, and both LV number of classes and re--estimate the level 1 measurement parameters.
						The second approach we describe is to conduct a ``full" 2.b by fixing only parameters in the multinomial logistic regression for $W$, and re--estimate $X \mid W$ and $Y \mid X$ multinomial logistic regression parameters for, say, $T = 1, \dots, T_{\text{max}}$.
						Although the latter approach is more computationally intensive, it could be useful in specific situations as described by \cite{Lukociene.10}: ``when the lower--level classes are poorly separated, information on the higher--level classes may help in finding the correct number of lower--level classes, provided  that  the  higher--level  classes  are  well  separated  themselves" \citep{Lukociene.10}. 
						
						One of the main working assumption of LC model parameter estimators is local independence of the indicators given the latent variable(s).
						We have assumed that this assumption holds in the IEA application.
						Local fit measures are available for multilevel latent class models \citep{nagelkerke2016,nagelkerke2017}, although the issue of detecting measurement non--invariance is not yet resolved as these are not asymptotically chi squared.
						Future research could focus on the two--stage estimator's behavior and adaptations to cases where the local independence assumption is violated (e.g., due to direct effects of covariates on the indicators).
							
							Another issue with two--step estimation for multilevel LC models is obtaining SE estimates that take into account uncertainty about the fixed parameters.
							Standard errors based on the Hessian of the last step are known to underestimate the true underlying variability of stepwise estimators \citep{zsuzsa2}.
							To obtain correct standard errors without increasing much the computational burden, we bootstrapped (non--parametrically) only stage 2 -- also in line with guidelines for stepwise LCA \citep{bakk2016}.
							In general, non parametric bootstrap can be a ``safe" option using available softwares.
							However, quicker options can be thought of, based on sandwich formulas \citep{white1982maximum} and correcting for underestimation of the true variability with additive corrections \citep{oberski}.
							Investigating theoretically correct SE estimation -- possibly with closed--form formulas -- is an important topic for future work.

							\appendix
							\section{Additional tables and figures}
							
							\begin{figure}[!h]
								\centering
								\includegraphics[width=\textwidth]{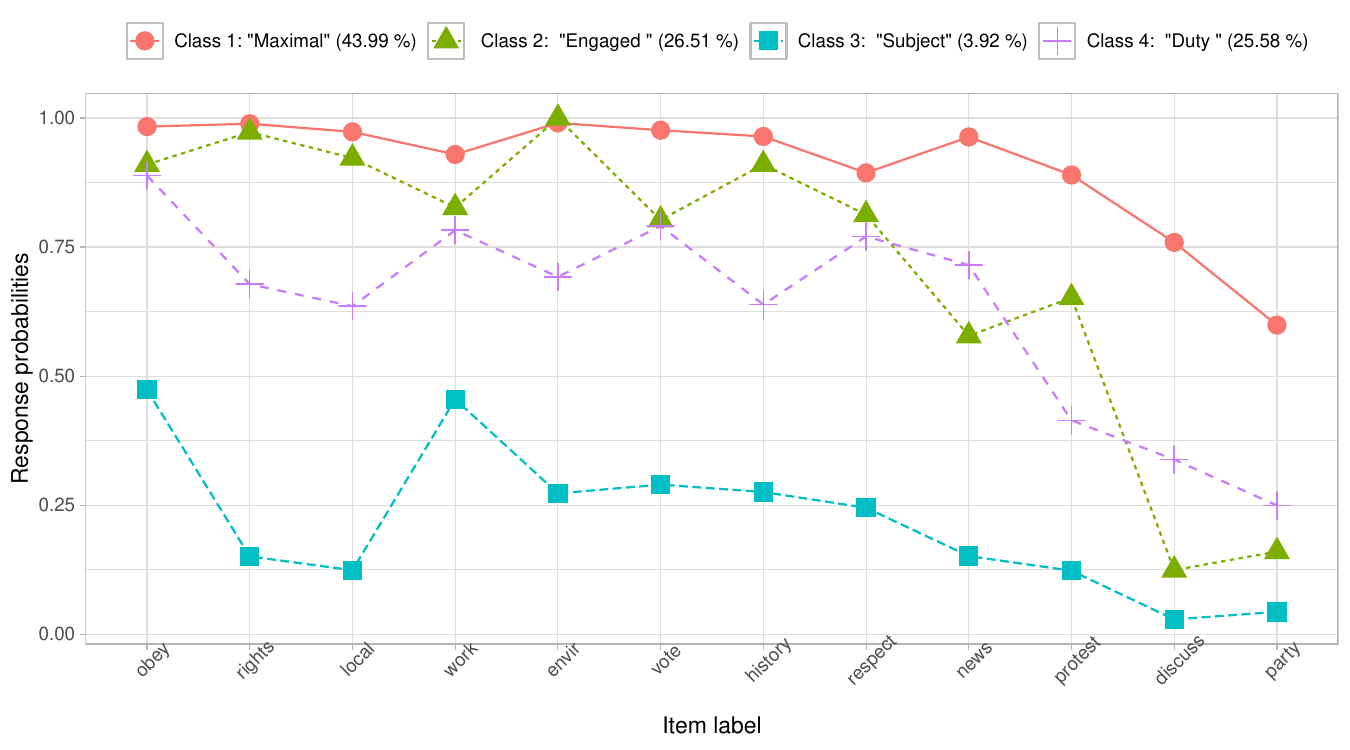}
								\caption{The basic latent class model (measurement model) at the lower level}
								\label{fig:Figbasic}
							\end{figure}

									\section{Model summaries}
									\begin{table}[!h]
										\begin{tabular}{lcccc}
											\hline \hline
											$T$	&	log--likelihood	&	BIC	&	Npar	&	entr. $R^2$	\\
											\hline
											1 &	-501351.0962	&	1002838.693	&	12	&	1	\\
											2 	&	-457351.5799	&	914987.5366	&	25	&	0.7213	\\
											3 	&	-448212.3953	&	896857.0434	&	38	&	0.6561	\\
											4 &	-444754.1647	&	890088.4583	&	51	&	0.6239	\\
											{\bf 5} &	{\bf -442942.6846}	&	{\bf 886613.3741}	&	{\bf 64}	&	{\bf 0.6039	}\\
											6 &	-441147.5251	&	883170.9312	&	77	&	0.6065	\\
											7 &	-440338.5716	&	881700.9001	&	90	&	0.5798	\\
											8 &	-439962.9764	&	881097.5857	&	103	&	0.5732	\\
											9 &	-439698.4559	&	880716.4206	&	116	&	0.5486	\\
											10 &	-439499.6612	&	880466.7072	&	129	&	0.5694	\\
											\hline
											\hline
										\end{tabular}
										\caption{Summary statistics for model selection at low level (Step 1). Selected $T$ in bold. \label{tab:modsum_step1}}
									\end{table}

									\begin{table}[!h]
										\begin{tabular}{lcccccc}
											\hline \hline
											$M$ 	&	log--likelihood	&	BIC	& BIC ($n_j$) & TIC &	Npar & entr. $R^2$ (low--lev)	\\
											\hline
											1  & -1702481.198 &3405007.867 &3404975.108 &3404962.415 &4 & 1 \\ 
											2  & -1701985.349 &3404073.007 &3403999.300 &3403969.852 &9 &0.792 \\ 
											{\bf 3}  & {\bf -1701799.232} & {\bf 3403757.612} & {\bf 3403642.957} & {\bf 3403597.627} & {\bf 14} & {\bf 0.814} \\ 
											4  & -1701985.579 &3404187.145 &3404031.541 &3403971.185 &19 &0.833 \\ 
											
											\hline
											\hline
										\end{tabular}
										\caption{Summary statistics for model selection at high level (Step 2a), with $nj=5$ as selected from Step 1. Selected $M$ in bold.
											\label{tab:modsum_step2a}}
									\end{table}

									\begin{table}[!h]
										\begin{tabular}{lcccc}
											\hline \hline
											$T$	&	log--likelihood	&	BIC	&	Npar	&	entr. $R^2$	\\
											\hline
											1	&	-484857.7962	&	969874.7406	&	14	&	1.0000	\\
											2	&	-446868.0265	&	894065.7169	&	29	&	0.6893	\\
											3	&	-436544.3193	&	873588.8183	&	44	&	0.6508	\\
											{\bf 4}	&	{\bf -432088.8293}	&	{\bf 864848.3541}	&	{\bf 59}	&	{\bf 0.6401}	\\
											5	&	-429566.1149 &	859973.4411	&	74	&	0.7136	\\
											6	&	-427043.1994	&	855098.1259	&	89	&	0.7534	\\
											7	&	-424477.0087	&	850136.2604	&	104	&	0.7415	\\
											8	&	-422633.8249	&	846620.4086	&	119	&	0.7387	\\
											9	&	-422010.3229	&	845543.9203	&	134	&	0.6991	\\
											10	&	-421339.4679	&	844372.7262	&	149	&	0.6917	\\
											\hline
											\hline
										\end{tabular}
										\caption{Summary statistics for model selection at low level (Step 2b), $M=3$. Selected $T$ in bold.
											\label{tab:modsum_step2b}}
									\end{table}

									%
									
									\FloatBarrier
									\section{Output from 5 and 6 low--level class solutions}
									
									\begin{figure}[!h]
										\centering
										\includegraphics[width=\textwidth]{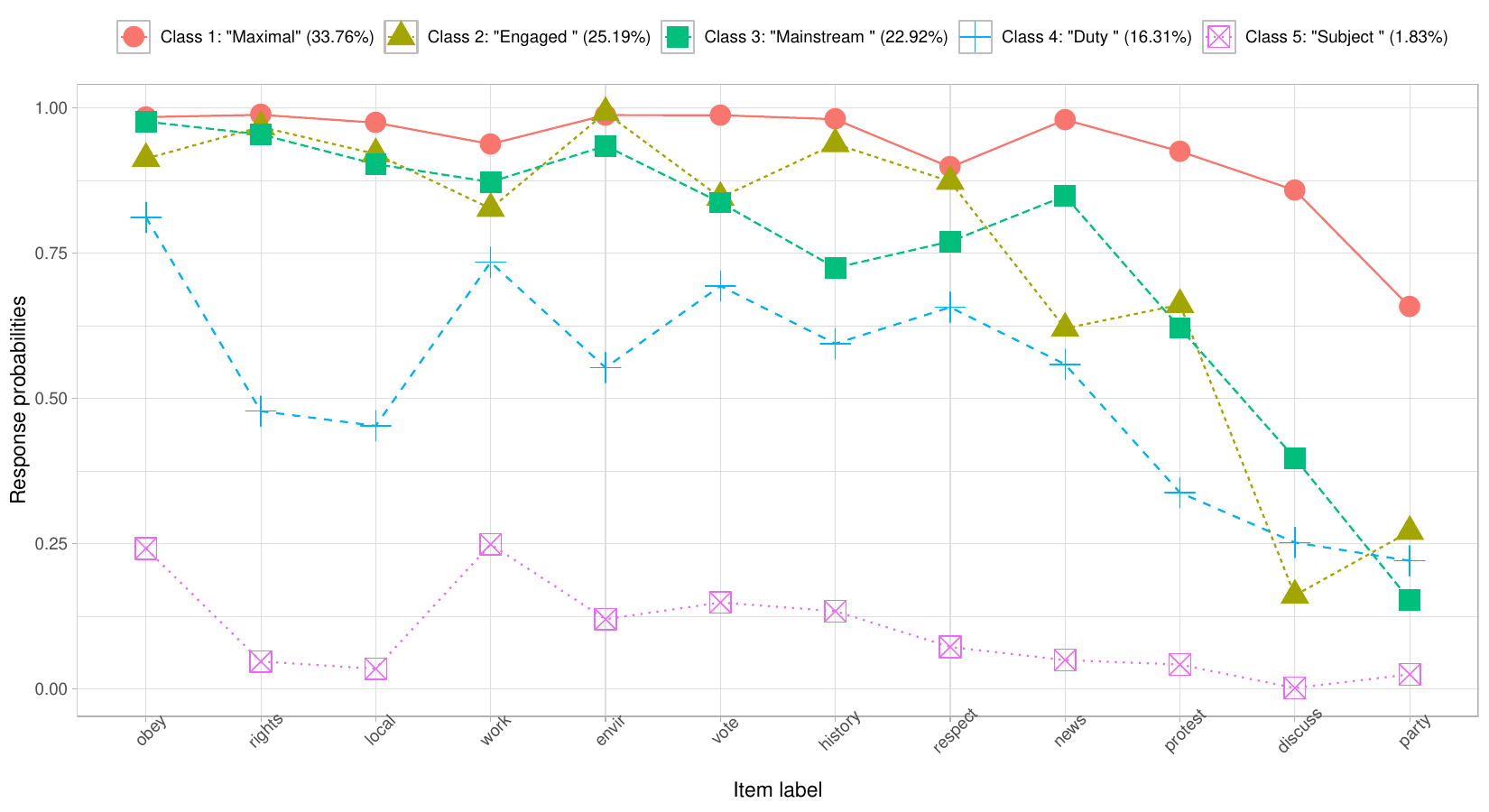}
										\caption{The basic latent class model (measurement model) at the lower level, with $T=5$, and $M=3$.}
										\label{fig:Figbasic_cl5}
									\end{figure}

									\begin{figure}[!h]
										\centering
										\includegraphics[width=\textwidth]{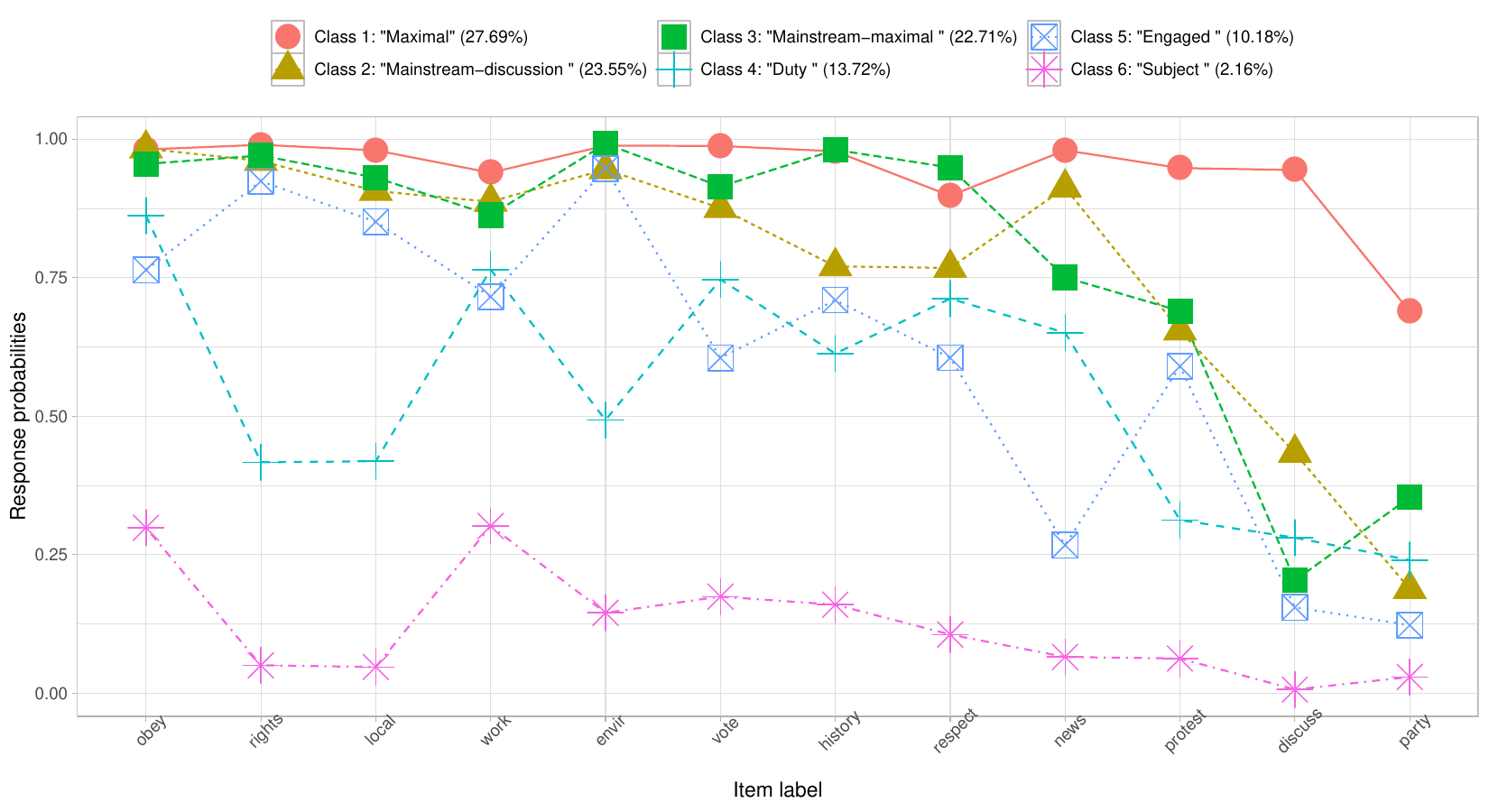}
										\caption{The basic latent class model (measurement model) at the lower level, with $T=6$, and $M=3$.}
										\label{fig:Figbasic_cl6}
									\end{figure}
									
									\begin{table}[!h]
											\begin{tabular}{l cccc}
												\hline \hline
												& Class 2 & Class 3 & Class 4 & Class 5 \\
												& (Engaged) & (Mainstream) & (Duty) & (Subject) \\
												\cmidrule{2-5}
												SES proxy &  -0.19$^{***}$ &  -0.10&  -0.23$^{***}$ &  -0.43$^{***}$\\ 
												& (0.05) & (0.08) & (0.05) & (0.11)\\ 
												Female &  0.40$^{***}$ &  0.30$^{***}$ &  -0.24$^{***}$ &  -0.43$^{**}$\\ 
												& (0.10) & (0.05) & (0.06) & (0.14)\\ 
												Education goal &  0.26 &  0.01 &  -0.34$^{***}$ &  -0.59$^{***}$\\ 
												& (0.15) & (0.14) & (0.06) & (0.10)\\ 
												Mother education &  0.05 &  -0.04 &  0.01 &  0.08\\ 
												& (0.04) & (0.08) & (0.04) & (0.10)\\ 
												Father education &  -0.2$^{*}$ &  -0.03 &  -0.13$^{**}$ &  -0.05\\ 
												& (0.08) & (0.05) & (0.04) & (0.10)\\ 
												Non-native language level &  -0.52$^{**}$ &  0.07 &  0.16 &  -0.28\\ 
												& (0.19) & (0.18) & (0.26) & (0.43)\\ 
												\hline \hline
											\end{tabular}
										\caption{Lower--level structural model parameter estimates for the two--stage estimator, with 5 lower level and 3 higher level classes. First class taken as reference. Standard errors, obtained with non parametric bootstrap on the 4th step, in parentheses. $^{***}$ $p$--value$<$0.01, $^{**}$ $p$--value$<$0.05, $^{*}$ $p$--value$<$0.1\label{table:reg_5cl}}
									\end{table}
									
									\begin{table}[!h]
										\resizebox{\textwidth}{!}{\begin{tabular}{l ccccc}
												\hline \hline
												& Class 2 & Class 3  & Class 4 & Class 5 & Class 6 \\
												& (Mainstream-discussion) & (Mainstream-maximal) & (Duty) &  (Engaged) & (Subject) \\
												\cmidrule{2-6}
												SES proxy &  -0.07 &  -0.16$^{***}$ &  -0.2$^{***}$ &  -0.33$^{*}$ &  -0.43$^{***}$\\ 
												& (0.08) & (0.03) & (0.04) & (0.13) & (0.1)\\ 
												Female &  0.37$^{***}$ &  0.5$^{***}$ &  -0.17$^{*}$ &  0.16 &  -0.38$^{**}$\\ 
												& (0.08) & (0.11) & (0.09) & (0.3) & (0.14)\\ 
												Education goal &  0.02 &  0.17 &  -0.37$^{**}$ &  0.15 &  -0.58$^{***}$\\ 
												& (0.15) & (0.23) & (0.11) & (0.31) & (0.14)\\ 
												Mother education &  -0.04 &  0.01 &  0.03 &  0.01 &  0.08\\ 
												& (0.11) & (0.07) & (0.06) & (0.11) & (0.11)\\ 
												Father education &  -0.02 &  -0.13 &  -0.09$^{***}$ &  -0.31 &  -0.07\\ 
												& (0.09) & (0.09) & (0.02) & (0.17) & (0.11)\\ 
												Non-native language level &  0.15 &  -0.59$^{*}$ &  0.25 &  -0.32 &  -0.28\\ 
												& (0.23) & (0.26) & (0.25) & (0.42) & (0.37)\\ 
												\hline \hline
										\end{tabular}}
										\caption{Lower--level structural model parameter estimates for the two--stage estimator, with 6 lower level and 3 higher level classes. First class taken as reference. Standard errors, obtained with non parametric bootstrap on the 4th step, in parentheses. $^{***}$ $p$--value$<$0.01, $^{**}$ $p$--value$<$0.05, $^{*}$ $p$--value$<$0.1\label{table:reg_6cl}}
									\end{table}
									
									\FloatBarrier
									\section{Survey questions}
									The survey question regarding citizenship norms is worded as follows: \citep{kohler2018} 
									
									\noindent Q23 How important are the following behaviors for being a good adult citizen? Respondent options: ``very important"; ``quite important"; ``not very important"; ``not important at all".
									
									\begin{table}[!h]
										\begin{tabular}{lll}
											\hline
											\hline
											IEA variable	&	Indicator	&	Questionnaire text	\\
											\hline
											IS3G23L	&	obey	&	Always obeying the law	\\
											&			&		\\
											IS3G23I	&	rights	&	Taking part in activities   \\  		    &			& promoting human rights	\\
											&			&		\\
											IS3G23H	&	local	&	Participating in activities to 	\\	        &           &  benefit people in\\
											&           & the local community	\\
											&			&		\\
											IS3G23K &	work	&	Working hard	\\
											&			&		\\
											IS3G23J	&	envir	&	Taking part in activities to \\     		& 			&protect the environment	\\
											&		&		\\
											IS3G23A	&	vote	&	Voting in every \\
											&           & national election	\\
											&		&		\\
											IS3G23C	&	history	&	Learning about the \\
											&			&  country's history	\\
											&			&		\\
											IS3G23E	&	respect	&	Showing respect for \\
											&			&	government representatives	\\
											&			&		\\
											IS3G23D	&	news	&	Following political issues \\
											&			&in the newspaper, on the \\
											&			& radio, on TV, or \\
											&			&	on the Internet	\\
											&			&		\\
											IS3G23G	&	protest	&	Participating in peaceful \\
											&			&	protests against laws \\
											&			&	believed to be unjust	\\
											&		&		\\
											IS3G23F	&	discuss	&	Engaging in political discussions	\\
											&		&		\\
											IS3G23B	&	party	&	Joining a political party	\\
											\hline \hline
										\end{tabular}
									\end{table}
									\FloatBarrier

									\newpage
									
									\bibliography{biblio}
									\bibliographystyle{apacite}
									
								\end{document}